%% file: wisedb-vldb16.tex
\documentclass{vldb}

\usepackage{latexsym}
\usepackage{times}
\usepackage{graphics}
\usepackage{color}
\usepackage{amssymb,amsmath,amsfonts}
\usepackage{wrapfig,epsfig}
\usepackage{subcaption}
\usepackage{color}
\usepackage{mathrsfs} 
\usepackage{xspace}
\usepackage{balance}
\usepackage{fancyvrb}
\usepackage[nocompress]{cite}

\newcommand{\Comment}[1]{\relax}

\newcommand{\Penalty}[2]{p(#1, #2)}
\newcommand{\Cost}[2]{cost(#1, #2)}
\newcommand{\cut}[1]{}
\newcommand{\XCloud}{WiSeDB\xspace}

\newtheorem{theorem}{Theorem}[section]
\newtheorem{lemma}[theorem]{Lemma}      
      
\def\compactify{\itemsep=0pt \topsep=0pt \partopsep=0pt \parsep=0pt}
\let\latexusecounter=\usecounter
\newenvironment{CompactEnumerate}
 {\def\usecounter{\compactify\latexusecounter}
  \begin{enumerate}}
 {\end{enumerate}\let\usecounter=\latexusecounter}

\begin{document}
\begin{sloppypar}

\title{WiSeDB: A Learning-based Workload Management Advisor for Cloud Databases}

\numberofauthors{2}
\author{
\alignauthor
Ryan Marcus \\
	\affaddr{Brandeis University} \\
	\email{ryan@cs.brandeis.edu}
\alignauthor
Olga Papaemmanouil \\
	\affaddr{Brandeis University} \\
	\email{olga@cs.brandeis.edu}
}

\maketitle

\begin{abstract}
Workload management for cloud databases deals with the tasks of resource provisioning, query placement, and query scheduling in a manner that meets the application's performance goals while minimizing the cost of using cloud resources. Existing solutions have approached these three challenges in isolation while aiming to optimize a single performance metric. In this paper, we introduce \XCloud, a learning-based framework for generating \emph{holistic} workload management solutions customized to application-defined performance goals and workload characteristics. Our approach relies on supervised learning to train cost-effective decision tree models for guiding query placement, scheduling, and resource provisioning decisions. Applications can use these models for both batch and online scheduling of incoming workloads.  A unique feature of our system is that it can adapt its offline model to stricter/looser performance goals with minimal re-training. This  allows us to present to the application alternative workload management strategies that address the typical performance vs. cost trade-off of cloud services. Experimental results show that our approach has very low training overhead while offering low cost strategies for a variety of performance metrics and workload characteristics.

\end{abstract}

\input{intro}


\input{system}


\input{problem_definition}

\input{model_generation}

\input{adaptive}

\input{runtime}

\input{experiments}

\input{related}
\balance

\input{conclusion}

\section{Acknowledgments}
This research was funded by NSF IIS 1253196.

\bibliographystyle{abbrv}
\bibliography{wisedb-vldb16}

\end{sloppypar}
\end{document}

%% file: intro.tex
\section{Introduction}\label{s:intro}

Cloud computing has transformed the way data-centric applications are deployed by reducing data processing services to commodities that can be acquired and paid for on-demand. Despite the increased adoption of cloud databases,
challenges related to workload management still exist, including provisioning cloud resources (e.g., virtual machines (VMs)), assigning incoming queries to provisioned VMs, and query scheduling within a VM in order to meet  performance goals. 
These tasks strongly depend on  application-specific workload characteristics and performance goals, and they are typically addressed by ad-hoc solutions at the application level.
 
 A number of efforts in cloud databases attempt to tackle these challenges (e.g., ~\cite{icbs,sla-tree,slos,cloudoptimizer,bazaar,smartsla,pmax,q-cop,activesla}).  However, these techniques suffer from two main limitations. First, they do not provide holistic solutions but instead address only individual aspects of the problem, such as  query admission~\cite{activesla, q-cop},  query placement to VMs~\cite{sla-tree, pmax, slos}, query scheduling within a VM~\cite{sla-tree, icbs}, or VM provisioning~\cite{slos, bazaar,cloudoptimizer,smartsla}. Since  these solutions are developed independently of each other, their integration into a unified framework requires substantial effort and investment to ``get it right'' for each specific case.
Second, while  a broad range of latency-related performance metrics are covered by these  systems (e.g., query response time ~\cite{sla-tree, smartsla, pmax, icbs}, average query latency~\cite{pmax}), each offers  solutions tuned only for a specific metric. Adapting them to support a wider range of application-specific metrics (e.g., max latency, percentile metrics) is not a straightforward task. 

Expanding on our vision~\cite{wisedb-clouddm}, we argue that cloud-based databases could benefit from a workload management advisor service that removes the burden of the above challenges from application developers. Applications should be able to specify their workload characteristics and performance objectives, and such a service should return a set of low-cost and performance-efficient strategies for executing their workloads on a cloud infrastructure. 

We have identified a number of design goals for such an advisor service. First, given an incoming query workload and a performance goal, the service  should provide \emph{holistic solutions} for executing a given  workload on a cloud database. Each solution should indicate: (a) the cloud resources  to be provisioned (e.g., number/type of VMs), (b) the distribution of resources among the workload queries (e.g., which VM will execute a given query), and (c) the execution order of these queries. We refer to these solutions collectively as {\em workload schedules}.

Second, to support diverse applications (e.g., scientific, financial, business-intelligence, etc), we need a customizable service that supports equally diverse application-defined performance criteria. We envision a \emph{customizable} service that generates workload schedules tuned to performance goals and workload characteristics specified by the application. Supported metrics should capture the performance of individual queries (e.g., query latency) as well as the performance of batch workloads (e.g., max query latency of an incoming query workload).

Third, since cloud providers offer resources for some cost (i.e., price/hour for renting a VM), optimizing schedules for this cost is vital for cloud-based applications. Hence, any workload management advisor should be \emph{cost-aware}. Cost functions are available through contracts between the service providers and their customers in the form of service level agreements (SLAs). These cost functions define the price for renting cloud resources, the performance goals, and the penalty to be paid if the agreed-upon performance is not met. A workload management service should consider \emph{all} these cost factors while assisting applications in exploring performance/cost trade-offs. Since different workload schedules offer different performance vs. cost trade-offs for different performance metrics, the system should be able to discover the ``best'' strategy for executing a given workload under an application-specific performance goal. Low cost workload schedules should be discovered independently of the performance metric.

This paper introduces \emph{\XCloud} ([W]orkload management [Se]rvice for cloud [DB]s), a workload management advisor for cloud databases designed to satisfy the above requirements. WiSeDB offers customized solutions to the workload management problem by recommending cost-effective strategies for executing incoming workloads for a given application. These strategies are expressed as \emph{decision-tree models} and \XCloud utilizes a supervised learning framework to ``learn''  models customized to the application's performance goals and workload characteristics.  For an incoming workload, \XCloud can parse the model to identify  the number/type of VMs to provision,  the  assignment of queries to VMs, and the execution order within each VM, in order to execute the workload and meet the performance objective with low-cost. 

Each model is cost-aware: it is trained on a set of performance and cost-related features collected from minimum cost schedules of  sample workloads. This cost  accounts for resource provisioning as well as any penalties paid due to failure to meet the performance goals. Furthermore, our proposed features are independent from the application's performance goal and workload specification, which allows WiSeDB to learn effective models for a range of metrics (e.g., average/max latency, percentile-based metrics). Finally, each model is trained offline once, and can be used at runtime to generate schedules for \emph{any} workload matching the model's workload specifications. Given an incoming batch workload and a decision model, \XCloud parses the model and returns a low cost  schedule for executing the  workload on cloud resources.

 \XCloud leverages a trained decision model in two additional ways. First, the training set of the  model is re-used to generate a set of \emph{alternative models} for the same workload specification, but stricter or more relaxed performance criteria. Several models are presented to the user, along with the cost estimate for each as function of the workload size. This allows applications to explore the performance vs. cost trade-off for their specific workloads.  Second, each model can be adjusted (with small overhead) during runtime to support \emph{online scheduling} of queries arriving one at a time. 

The contributions of this work can be summarized as follows:

\begin{CompactEnumerate}
\item We introduce {\em \XCloud}, a workload management advisor for cloud databases. We discuss its  design and rich functionality, ranging from recommendations of alternative workload execution strategies that enable exploration of performance vs. cost trade-offs to resource provisioning and query scheduling for both batch and online processing. All recommendations are tuned for application-defined workload characteristics and (query latency-related) performance goals.   

\item We propose a novel \emph{learning approach to the workload management problem}. WiSeDB learns its custom strategies by collecting features from optimal (minimum cost) schedules of sample workloads. We rely on a graph-search technique to identify these schedules and generate a training set in a timely fashion. Furthermore, we have identified a set of  performance and cost-related features that can be easily extracted from these optimal solutions and allow us to learn effective (low-cost) strategies for executing any workload for various performance metrics.

\item We propose an {\em adaptive} modeling technique that generates alternative workload execution strategies, allowing applications to explore  performance vs. cost trade-offs. 

\item We leverage the trained models to schedule batch workloads as well as to support online scheduling by efficiently generating low-cost models upon arrival of a new query.

\item We discuss experiments that demonstrate \XCloud's ability to learn low-cost strategies for a number of performance metrics with very small training overhead. These strategies offer effective solutions for both batch and online scheduling independently of workload and performance specifications. 
\end{CompactEnumerate}

We first discuss WiSeDB's system model in Section~\ref{s:system} and then define our optimization problem in Section~\ref{s:problem}. We introduce our modeling framework in Section~\ref{s:model} and the adaptive modeling approach in Section~\ref{sec:shift}. Section~\ref{s:runtime} discusses \XCloud's runtime functionality. Section~\ref{sec:expr} includes our experimental results. Section~\ref{s:related} discusses related work and we conclude in Section~\ref{s:conclusions}.


%% file: system.tex
\section{System Model}\label{s:system}

Our system is designed for data management applications deployed on an Infrastructure-as-a-Service (IaaS) cloud (e.g.,~\cite{url-amazonAWS,url-msazure}). These providers typically offer virtual machines (VMs) of different types (i.e., resource configurations) for a certain fee per renting period. We assume this deployment is realized by renting VMs with preloaded database engines (e.g., PostgreSQL~\cite{url-postgres}) and that queries can be executed locally on any of the rented VMs. This property is  offered by fully replicated databases.\footnote{\small Partial replication/data partitioning models can also supported by specifying which data partitions can serve a given query.}

\begin{figure}
	\centering 
	 \includegraphics[width=0.4\textwidth] {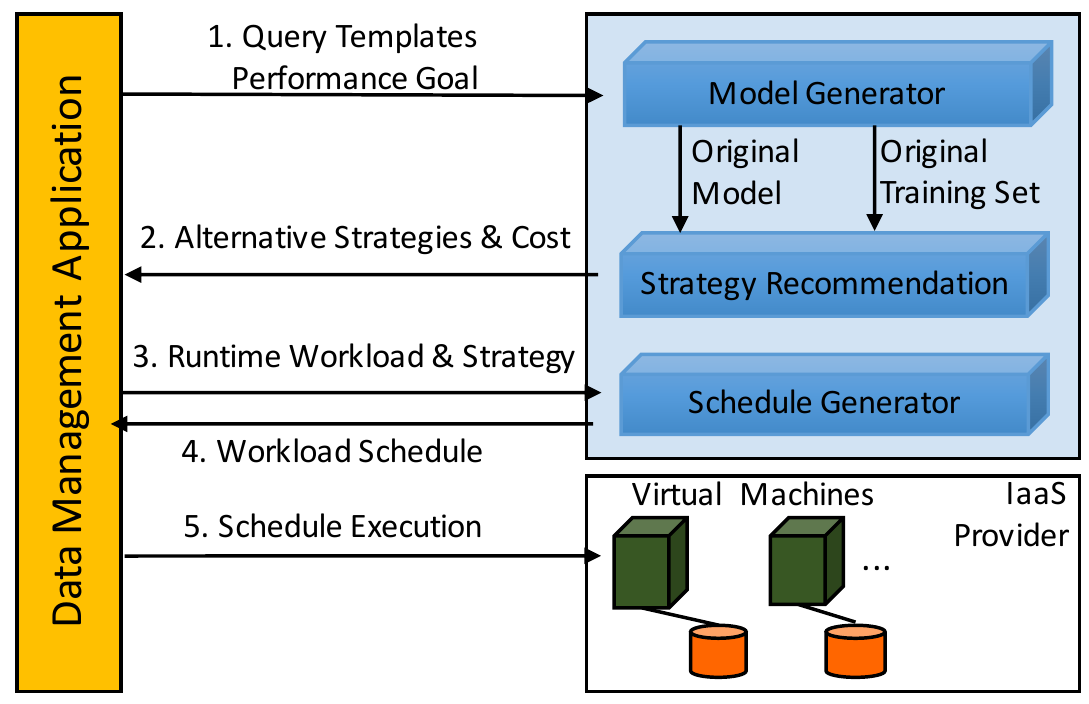}
	\caption{\small{The \XCloud system model} }
	\label{fig:system-mod}
	\vspace{-0.5em}
\end{figure}

{\bf Workload Specification} Applications begin their interaction with \XCloud by providing a \emph{workload specification} indicating the query templates (e.g., TPC-H templates~\cite{url-tpch})  that will compose their workloads. In analytical applications, incoming queries are generally instances of a small number of templates, and queries of the same template have similar performance properties (i.e.,  latency) because they access and join the same tables~\cite{tpch}. In practice, \XCloud's approach is agnostic to the tables accessed or joined {and cares only about the latency of each template, i.e., queries with identical latency can be treated as instances of the same template.} \ 

\begin{figure}
\centering
\begin{Verbatim}[frame=single]
SELECT  SUM (l_extendedprice * l_discount) 
FROM    lineitem
WHERE   l_shipdate >= date'[DATE]' 
AND     l_shipdate < date'[DATE]' 
        + interval'1'year 
AND     l_discount between [DISC] - 0.01 
        and [DISC] + 0.01 
AND     l_quantity < [QUANT];
\end{Verbatim}
\vspace{-4mm}
\caption{TPC-H Query Template \#6}
\label{fig:tpch}
\vspace{-3mm}
\end{figure}

{\bf Query Templates} We assume that a user's workload is composed of queries, each of which are derived from a finite set of \emph{query templates}, much like TPC-H templates~\cite{url-tpch}. A query template, generally expressed in SQL, is a query with a set of missing values. Users can instantiate a new query of a given template by providing the missing values. 
Figure~\ref{fig:tpch} shows an example of a query template from the TPC-H benchmarking suite. A new instance of this query template could be created by giving values for \texttt{[DATE]}, \texttt{[DISC]}, and \texttt{[QUANT]}. Conversely, any query matching the semantics of some template for a certain set of values can be said to be an instance of that template. 

{\bf Performance Goals} Applications also specify their performance goals for their  workloads as functions of query latency. Performance goals can be defined either at the query template level or workload level. Currently, we support the following four types of metrics. (1) {\em Per Query Deadline}: users can specify an upper latency bound for each query template (i.e., queries of the same template have the same deadline). (2) \emph{Max Latency Deadline}: the user expresses an upper bound on the worst query response time in a  query workload. (3) {\em Average Deadline}: sets an upper limit on the average query latency of a  workload. (4) {\em Percentile Deadline}: specifies that at  least $x\%$ of the workload's queries must be completed within $t$ seconds. These metrics cover a range of performance goals typically used for database applications {(e.g.,~\cite{icbs,pmax,q-cop,activesla})}. However, \XCloud is the first system to support them within a single workload management framework. 

Performance goals are expressed as part of a Service-Level-Agreement (SLA) between the IaaS provider and the application that states (a) the workload specification, (b) the performance goal, and (c) a penalty function that defines the penalty to be paid to the application if that goal is not met. Our system is agnostic to the details of the penalty function, {incorporating it into its cost-model as a ``black box'' function that maps performance to a penalty amount.}

The architecture of the WiSeDB Advisor is shown in Figure~\ref{fig:system-mod}. Workload and performance specifications are submitted to \XCloud, which trains a decision model,  a.k.a. \emph{strategy} (\emph{Model Generator}). The training set for this model is also leveraged to generate alternative decision models (strategies) for stricter and more relaxed performance goals ({\em Strategy Recommendation}). These strategies are presented to the user along with a cost function that estimates the monetary cost of each strategy based on the frequency of each query template in a given workload.

Given an incoming workload at runtime, the application estimates the expected cost and performance of executing these workloads using our proposed strategies and chooses the one that better balances performance needs and budget constraints (\emph{Execution Strategy}). WiSeDB identifies (a) the type/number of VMs to be provisioned, (b) the assignment of queries  to these VMs and (c) the execution order of the queries within each VM, in order to to execute the  incoming workload based on the chosen strategy.  This step is executed by the {\em Schedule Generator} which can generate these {\em workload schedules} for both batch workloads  as well as single queries as they arrive (online scheduling).  Applications  execute their workloads according to \XCloud's recommendations. They rent VMs as needed and add queries to the processing queue of VMs according to the proposed schedule. VMs are released upon completion of the workload's execution. In the case of online processing, the application adds the new query to the processing queue of an existing VM or to a new VM.

%% file: problem_definition.tex
\section{Problem definition}\label{s:problem}

Here, we formally define our system's optimization goal and discuss the problem's complexity. Applications provide a set of \textit{query templates} $T= \{T_1, T_2, \dots\}$ as the workload specification and a \textit{performance goal} $R$. Given the set of query templates $T$, \XCloud generates decision models for scheduling workloads with queries instances drawn from $T$. Let us assume a workload $Q=\{q_1^x, q_2^y, \dots\}$ where each query $q_i^j \in Q$ is an instance of the template $T_j \in T$. Given a workload $Q$ and one of the generated decision models, \XCloud identifies a  \emph{schedule} $S$  for executing $Q$.

We  represent a VM of type $i$, $vm^i$, as a queue $vm^i=[q_1^x, q_2^y, \dots]$ of queries to process in that order and we represent a \emph{schedule} $S=\{vm^i_1, vm^j_2, \dots \}$ of the workload $Q$ as a list of VMs such that each VM contains only queries from $Q$. Hence, each schedule $S$ indicates (1) the number and type of VMs to be provisioned,  (2) the assignment of each query $q^i_j \in Q$ to these VMs and (3) the query execution order on each VM, $vm^i_j \in S$. A \emph{complete} schedule assigns each query in $Q$ to one VM.

We denote the latency of a query $q^x_j$ (of template $T_x$) when executed on a VM of type $i$, $vm^i_k$, as $l(q^x_j, i)$. Latency estimates can be provided by either the application (e.g., by executing representative queries a-priori on the available VM types) or by using existing prediction models (e.g.,~\cite{jennie_sigmod11, contender}). Queries assigned to a VM can be executed immediately or can be placed in a VM's processing queue if no more concurrent queries can be processed.\footnote{\small Most DBMSs put an upper limit on the number of concurrent queries, referred to as multiprogramming level.}

\begin{figure}
\centering
\includegraphics[width=0.45\textwidth]{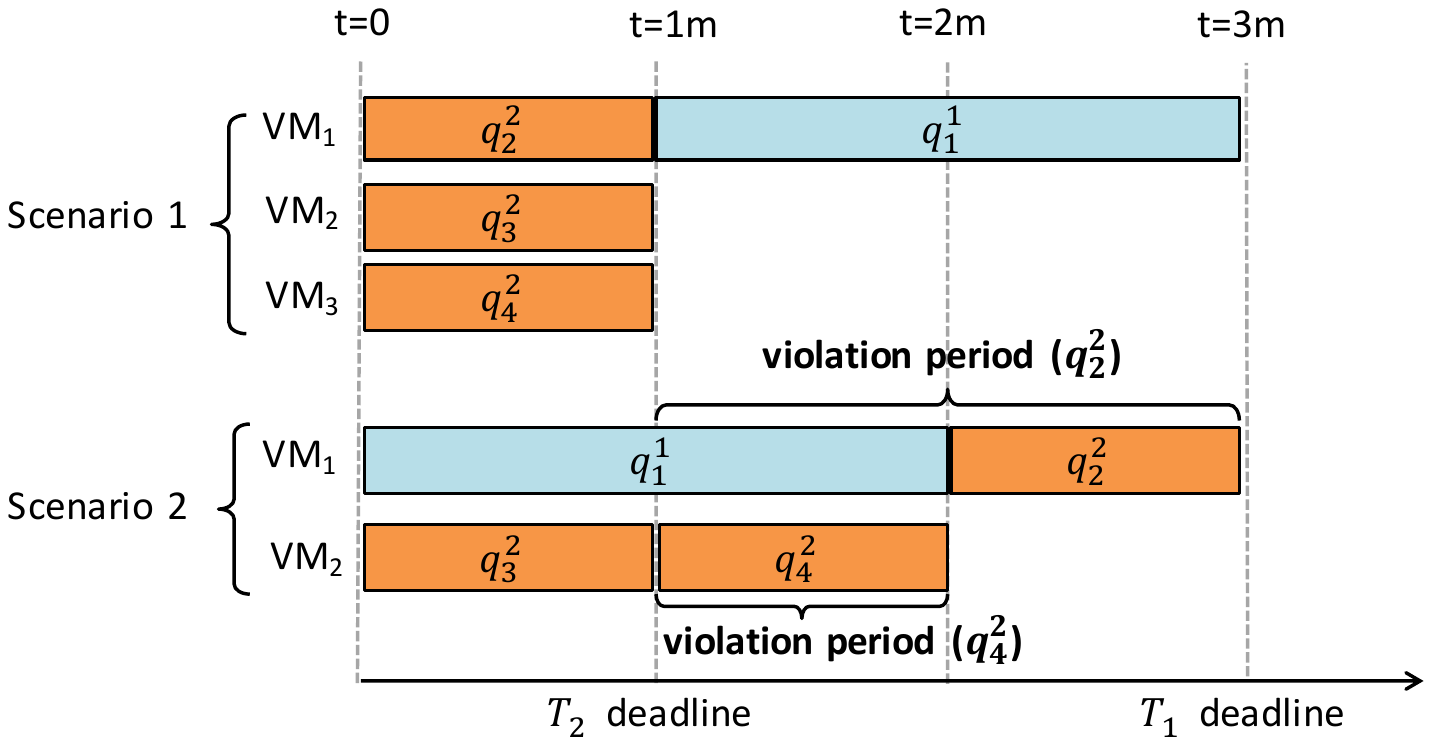}
\vspace{-1.0mm}
\caption{\small Two different schedules for $Q=\{q^1_1, q^2_2, q^2_3, q^2_4\}$} 
\label{fig:example_sla}
\vspace{-4.6mm}
\end{figure}

Figure~\ref{fig:example_sla} shows two possible schedules for a workload of four queries drawn out of two templates. The first scenario uses three VMs while the second executes the queries on two VMs. Based on our notation, the first schedule is expressed as $S_1 = \{ vm_1=[q_2^2, q_1^1], vm_2=[q^2_3], vm_3=[q^2_4] \}$, where as the second scenario is expressed as $ S_2 = \{ vm_1 = [q_1^1, q_2^2], vm_2 = [q^2_3, q^2_4] \}$. 
Here, $\Penalty{R}{S_2} \neq 0$, since the violation period for the schedule represented in the second scenario is not zero.

{\bf Cost Model} To calculate the monetary cost of processing a workload, we assume each VM of type $i$ has a fixed start-up cost $f^i_s$, as well as a running cost $f^i_r$ per unit of time (i.e. the price for renting the VM for that time unit). We also assume a penalty function $\Penalty{R}{S}$ that estimates the penalty for a given schedule $S$ and performance goal $R$. Without loss of generality (and similarly to the model used by IaaS providers~\cite{url-amazonAWS}), we assume penalties are calculated based on the violation period, i.e., a fixed amount per time period of violation within a schedule $S$.

The \emph{violation period} is the duration of time that the performance goal $R$ was not met. If the goal is to complete queries of a given template within a certain deadline (per query deadline goal), the violation period for each query is measured from the time point it missed its deadline until its completion. Figure~\ref{fig:example_sla} shows an example of this case. Let us assume that queries of template $T_1$ have an execution time of $2$ minutes and queries of $T_2$ execute with $1$ minute. The figure shows the violation periods for the second scenario (for $q^2_2, q^2_4)$ assuming that the deadline of template $T_1$ is $3$ minute and for $T_2$ is $1$ minutes (the first scenario does not have any violations).  

For the maximum latency metric, where no query can exceed the maximum latency, the violation period is computed in the same way. For an average latency performance goal, the duration of the violation period is the difference between the desired average latency and the  actual average latency of each query. For a percentile performance goal that requires $x\%$ of queries to be complete in $y$ minutes, the violation period is the amount of time in which $100-x\%$ of the queries had latencies exceeding $y$ minutes. We denote the monetary cost of performance violations for a given performance goal $R$ and a schedule $S$ as $\Penalty{R}{S}$.

{\bf Problem Definition} Given a workload $Q=\{q_1^x, q_2^y, \dots\}$, where $q_i^j$ is of template $T_j$, and a performance goal $R$, our goal is to  find a complete schedule $S$ that minimizes the total monetary cost (provisioning, processing, and penalty payouts costs) of executing the workload. We define this total cost, $\Cost{R}{S}$, as:
\begin{equation}
\label{eq:cost}
\Cost{R}{S} = \sum_{vm^i_j \in S} \left[ f^i_s +\sum_{q^m_k \in vm^i_j} f^i_r \times l(q^m_k, i) \right] + \Penalty{R}{S}
\end{equation}
{\bf Problem Complexity} Under certain conditions, our optimization problem becomes the bin packing problem, where we try to ``pack'' each query  into one of the available ``VM-bins''. For this reduction, we need to assume that (1) the number of query templates is unbounded, (2) infinite penalty, $\Penalty{R}{S} = \infty$, if the performance goal is violated, and (3) the start-up cost $f^i_s$ is uniform across all VM types. Under these conditions, the problem is {\tt NP-Hard}. However, these assumptions are not valid in our system. Limiting the number of query templates relaxes the problem to one of polynomial complexity, but still not computationally feasible~\cite{appalg}.

\begin{table}
\centering
\begin{tabular}{| r | l |}
\hline
\bf Symbol & \bf Meaning \\
 \hline
$R$              & performance goal \\
$S$              & workload schedule \\
$T_i$            & query template $i$ \\
$q^i_j$          & $j$-th query (of template $i$) \\
$f^i_s$          & start-up cost of VM type $i$ \\
$f^i_r$          & cost of renting VM type $i$ per unit time\\
$l(q^j, i)$      & latency of query template $j$ on VM type $i$\\
$\Penalty{R}{S}$ & penalty of $S$ under $R$\\
$\Cost{R}{S}$    & total cost of $S$ under $R$\\
$vm^i_j$         & $j$th VM (of type $i$) \\
$vm^i$           &  VM of type $i$ \\
$v_s$            & partial schedule at vertex $v$\\
$v_u$            & unassigned queries at vertex $v$\\
\hline
\end{tabular}
\vspace{-0.5em}
\caption{Notation table}
\vspace{-4mm}
\label{tbl:symbols}
\end{table}

Two common greedy approximations to this optimization problem are the first-fit decreasing (FFD)~\cite{ffd} and first-fit-increasing (FFI) algorithms, which sort queries in decreasing or increasing order of latency respectively and place each query on the first VM where the query ``fits'' (incurs no penalty). If the query will not fit on any VM, a new VM is created. Existing cloud management systems have used FFD (e.g.,~\cite{sci_place}) for provisioning resources and scheduling queries. However, it is often not clear which of these greedy approaches is the best for a specific workload and performance goal. For example, when applied to the workload and performance goals shown in Figure~\ref{fig:example_sla}, FFD schedules all queries on their own VM, which offers the same performance as scenario 1 but uses an additional VM (and hence has higher cost). A better approximation would be FFI, which produces the schedule that is depicted in scenario 1, scheduling the four queries across three VMs without violating the performance goal.

Furthermore, there might be scenarios when none of these approximations offer the best solution. For example, consider workloads consisting of three templates, $T_1, T_2, T_3$ with latencies of four, three, and two minutes respectively. Assume we have two queries of each template, $q^1_1, q^1_2$ of template $T_1$, $q^2_3, q^2_4$ of template $T_2$, and $q^3_5, q^3_6$ of template $T_3$ and we wish to keep the total execution time of the workload below nine minutes. FFD will find the schedule $S_{FFD} = \{[q^1_1, q^1_2], [q^2_3, q^2_4, q^3_5], [q^3_6]\}$, while FFI would find the schedule $S_{FFI} = \{q^3_5, q^3_6, q^2_3], [q^2_4, q^1_1], [q^1_2]\}$. However, a better strategy is one that attempts to place an instance of $T_1$, then an instance of $T_2$, then an instance of $T_3$, then create a new VM, resulting in the schedule $S^\prime = \{[q^1_1, q^2_3, q^3_5], [q^1_2, q^2_4, q^3_6]\}$, which has a lower cost because it provisions one less VM.

\emph{\XCloud departs from the ``one-strategy-fits-all'' approach used by standard approximation heuristics. Instead, it offers effective scheduling strategies for custom performance goals and workload specifications by learning heuristics tailored  to the application's needs.}
We next describe how \XCloud identifies such strategies.

%% file: model_generation.tex
\section{Decision Model Generation}\label{s:model}
\XCloud relies on supervised learning to address our workload management problem. Next, we describe this process in detail.

\subsection{Approach Overview}

Given an application-defined workload specification (i.e., query templates  and a performance goal), WiSeDB generates a set of workload execution strategies that can be used to execute incoming query workloads with low cost and within the application's performance goal. Formally, our goal is to identify strategies that minimize the total cost as defined in Equation~\ref{eq:cost}. Towards this end, our framework generates samples of optimal schedules (i.e., that minimize the total cost) and relies  on decision tree classifiers to learn ``good'' strategies from these optimal solutions.

\begin{figure}[t]
\centering
\includegraphics[width=0.48\textwidth]{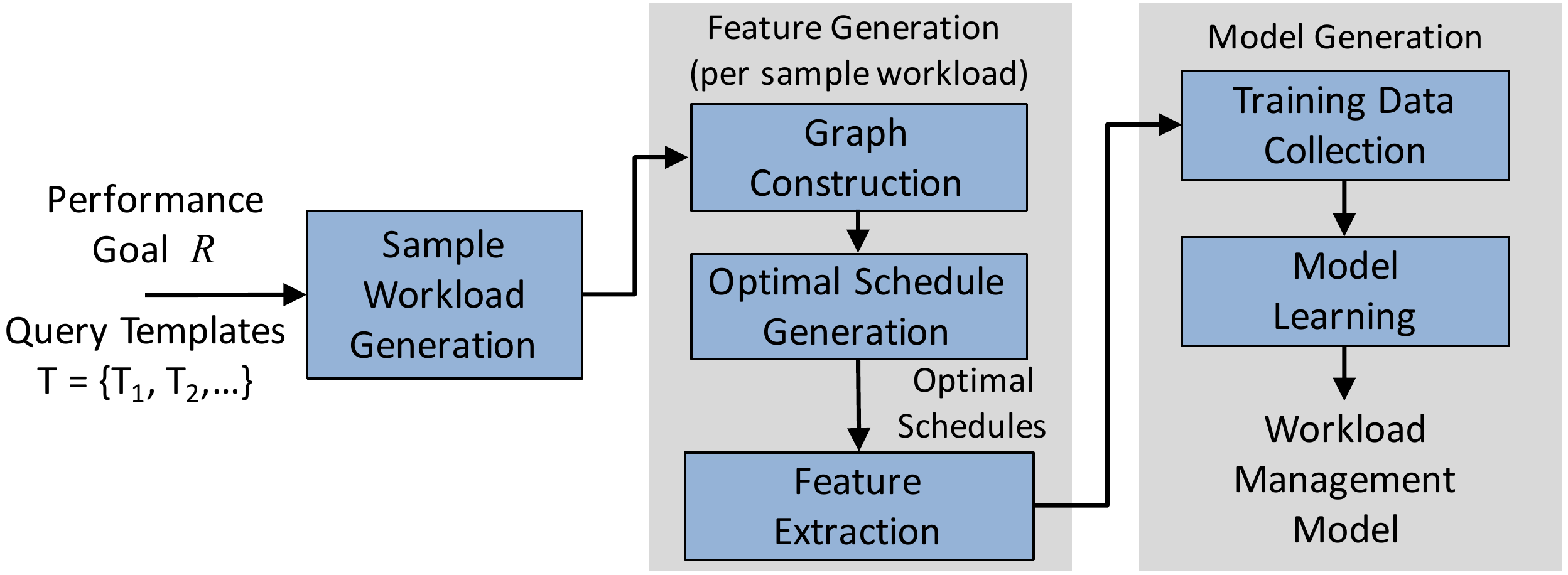}
\vspace{-5.9mm}
\caption{\small{Generation of the decision model}}
\vspace{-4.9mm}
\label{f:training}
\end{figure}

Our framework is depicted in Figure~\ref{f:training}.  Initially, we create a large number of \emph{random sample workloads}, each consisting of a small number of queries drawn from the query template definitions. Our next step is to identify the optimal schedule for each of these sample workloads. To do this efficiently, we represent the problem of scheduling workloads as a \emph{graph navigation} problem. On this graph, edges represent query assignment or resource provisioning decisions and the weight of each edge is equal to the cost of that decision. Hence, each path through the graph represents decisions that compose some schedule for the given workload. Finding the optimal schedule for that workload is thus reduced to finding the shortest path on this graph. 

Next, for each decision within an optimal schedule, we extract a set of features that characterize this decision. We then generate a  training set which includes all  collected features from all the optimal schedules across all sample workloads. Finally, we train a \emph{decision tree model} on this training set. The learning process is executed offline and the generated models can be used during runtime on incoming workloads. Next, we describe these steps in detail.

\subsection{Workload Sampling}

\XCloud first generates sample workloads based on the application-provided query templates $T$. We create $N$ random sample workloads, each containing $m$ queries. Here, $N$  and $m$ must be sufficiently large so that query interaction patterns emerge and the decision tree model can be properly trained.  However, $m$ must also be sufficiently small so that for each sample workload we can identify the optimal schedule in a timely manner. 

In order to ensure that our sampling covers the space of possible workloads, we rely on uniform direct sampling of the query templates. If the sampling is not uniform, the decision tree may not be able to learn how two query templates interact, or the decision tree may have very little information about certain templates. Furthermore, we generate a large number of samples
(e.g., in our experiments $N=3000$) 
in order to ensure that our workload samples will  also include workloads that are imbalanced with respect to the number of unique templates they include. This allows WiSeDB to handle skewed workloads.


Even though our training will be based on uniform sampling of the query templates, \XCloud can still handle skewed workloads. {This is because uniform direct sampling generates some sample workloads that are balanced and some sample workloads that are not balanced.  The unbalanced sample workloads teach the decision tree classifier how to handle odd, skewed workloads, and the balanced workloads teach the decision tree how to handle the ``usual'' mixes.} We demonstrate this experimentally in Section~\ref{s_sensitivity}.

\subsection{Optimal Schedule Generation}\label{sec:graph}

Given a set of sample workloads, WiSeDB learns a model based on the optimal schedules for these workloads. To produce these optimal solutions, we represent schedules as paths on a weighted graph, and we find the minimum cost path on this graph. This graph-based approach provides a number of advantages. First, each ``best'' path represents not only an optimal schedule, but also \emph{the steps taken to reach that optimal schedule}. The information relevant to each optimal decision is captured by each vertex's state.  \cut{Second, the highly-structured nature of these graphs enable us to \emph{isolate all the information relevant to the optimal decision} within a vertex.} Second, a graph representation lends itself to a natural way of \emph{eliminating redundancy in the search space} via careful, strategic path pruning. Finally, the well-studied nature of \emph{shortest-path problems} enables the application of deeply-understood algorithms with desirable properties. Next, we describe our graph construction in detail, and highlight these advantages.

{\bf Graph Construction} Given a sample workload $Q=\{q_1^x, q_2^y, \dots\}$, we construct a directed, weighted graph $G(V,E)$ where vertices represent intermediate steps in schedule generation, i.e., partial schedules and a set of remaining queries to be scheduled. Edges represent actions, such as renting a new VM or assigning a query to a VM. The cost (weight) of each edge will be the cost of performing a particular action (e.g., the cost of starting a new VM). We refer to this as a \emph{scheduling graph}. 

\begin{figure}[t]
\centering
\includegraphics[width=0.45\textwidth]{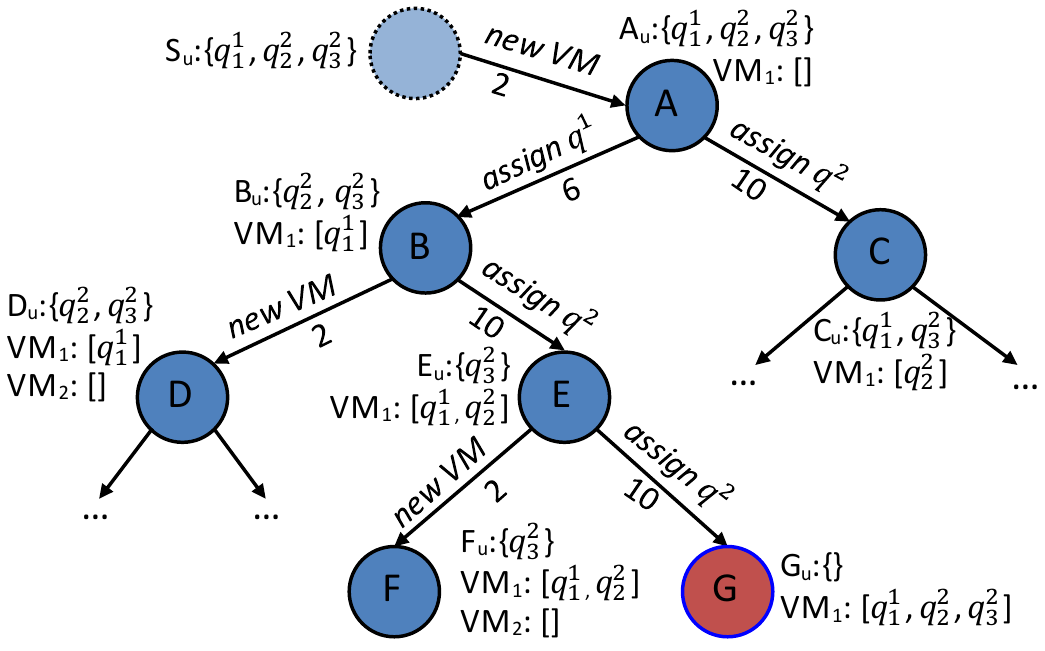}
\vspace{-0.45em}
\caption{ \small{A subgraph of a scheduling graph for two query templates and $Q=\{q^1_1, q^2_2, q^2_3\}$. $G$ is one goal vertex}}
\vspace{-3.5mm}
\label{fig:graph}
\end{figure}

Formally, each vertex $v\in V$  has a schedule for some queries of the workload, $v_s=\{vm^i_1, vm^k_2,\dots\}$, which includes the VMs to be rented for these queries. Each VM $j$ of type $i$, $vm^i_j$, is a queue of queries that will be processed on that VM, $vm^i_j=[q^x_k, q^y_m,...]$. Hence, $v_s$ represents possible (potentially partial) schedules for the workload $Q$. Each $v$ also has a set of unassigned queries from $Q$, $v_u$, that must still be placed onto VMs. 

The \emph{start vertex}, $A \in V$, represents the initial state where all queries are unassigned. Therefore, $A_u$ includes all the queries in the given workload and $A_s$ is empty. If a vertex $g \in V$ has no unassigned queries, we say that vertex $g$ is a \emph{goal vertex} and its schedule $g_s$ is a \emph{complete} schedule. 

An edge in $E$ represents one of two possible actions:

\begin{CompactEnumerate}
\item{A \emph{start-up edge} $(u, v, i)$ represents renting a new VM of type $i$, $vm^i_j$. It connects a vertex $u$ to $v$ where $v$ has an additional empty VM, i.e., $v_s = u_s \cup vm^i_j$. It does not assign any queries, so $u_u = v_u$. The weight of a start-up edge is the cost to provision a new VM of type $i$: $w(u, v) = f^i_s$.}

\item{A \emph{placement edge} $(u, v, q^x_y)$ represents placing an unassigned query $q^x_y \in u_u$ into the queue of a rented VM in $v_s$. It follows that $v_u = u_u - q^x_y$. {Because \XCloud is agnostic to the specifics of any particular query, the placement of an instance of query template $T_x$ is equivalent to placing any other instance of $T_x$.} Therefore, we include only a single placement edge per query template even if the template appears more than once in $u_u$. The cost of an edge that places query $q^x_y$ into a VM of type $i$ is the execution time of the query multiplied by the cost per unit time of the VM, plus any \emph{additionally} incurred penalties: }
\begin{equation}
\label{eq:weight}
w(u, v) = \left[ l(q^x_y, i) \times f^i_r \right] + \left[\Penalty{R}{v_s} - \Penalty{R}{u_s}\right]
\end{equation}

\end{CompactEnumerate}

Figure~\ref{fig:graph} shows part of a scheduling graph for a workload $Q=\{q_1^1,q^2_2, q^2_3\}$. $A$ represents the start vertex and $G$ represents a goal vertex.\footnote{\scriptsize{The dotted vertex represents the provisioning of the first VM which is always the first decision in any schedule.}} At each edge, a query is assigned to an existing VM ($\overline{AB}$, $\overline{AC}$ ,$\overline{BE}$, $\overline{EG}$), or a new VM is created ($\overline{BD}$, $\overline{EF}$). The path $\overline{ABEG}$ represents assigning the three queries, $q_1^1,q^2_2, q^2_3$, to be executed in that order on $vm_1$.

The weight of a path from the start vertex $A$ to a goal vertex $g$ will be equal to the cost of the complete schedule of the goal vertex, $\Cost{R}{g_s}$, for a given performance goal $R$. Since all complete schedules are represented by some goal state, searching for a minimum cost path from the start vertex $A$ to any  goal vertex $g$ will provide an optimal schedule for the workload $Q$. We then find the optimal path through the graph, noting the scheduling decision made at every step, i.e., which edge was selected at each vertex.


{\bf Graph Reduction} To improve the runtime of the search algorithm, we reduce the graph in a number of ways. {First, we include a start-up edge only if the last VM provisioned has some queries assigned to it, i.e.,  we allow renting a new VM \emph{only if the most recently provisioned VM is not empty}. {This eliminates paths that provision VMs that are never used.} Second, queries are assigned only to the most recently provisioned VM, i.e., each vertex has outgoing placement edges that assign a query \emph{only to the most recently added VM}.} {This reduces the number of redundant paths in the graph, since each {combination of VM types and query orderings} is accessible by only a single instead of many paths.} This reduction can be applied without loss of optimality, e.g. without eliminating any goal vertices. 

\begin{lemma}
Given a scheduling graph $G$ and a reduced scheduling graph $G_r$ all goal vertices with no empty VMs in $G$ are reachable in $G_r$.
\end{lemma}

\begin{proof}
Consider an arbitrary goal vertex with no empty VMs $g \in G$. For any vertex $v$, let $head(v_s)$ be the most recently created VM in $v_s$ (the head of the queue).

Let us assume $head(g_s) = vm^i_j$ and $q^x_y$ is the last query scheduled in $vm^i_j$. Then there is a placement edge $e_p = (v, g, q^x_y)$ connecting some vertex $v$ with query $q^x_y$ in its set of unassigned queries, i.e., $q^x_y \in v_u$, to $g$. Further, we know that $e_p$ is an edge of $G_r$ because $e_p$ is an assignment of a query to the most recently created VM. If $head(v_s)$ is non-empty, and $q^m_k$ is the last query on $head(v_s$), then there is another vertex $u$ connected to $v$ via a placement edge $e_p = (u, v, q^m_k)$ in $G_r$ by the same argument. If $head(v_s) = vm^i_j$ is empty, then there is a start-up edge $e_s = (y, v, i)$ connecting some vertex $y$, with the same set of unassigned queries as $v$, i.e., $v_u = y_u$, to $v$. We know that $e_s$ must be an edge of $G_r$ because $v$ can have at most one empty VM. This process can be repeated until the start vertex is reached. Therefore, there is a path from the start vertex in $G_r$ to any goal vertex with no empty VMs $g \in G$.
\end{proof}

{\bf Search Heuristic} WiSeDB searches for the minimum cost path (i.e., optimal schedule) from the start vertex to a goal vertex using the A* search algorithm~\cite{astar} which offers a number of advantages. First, it is {complete}, meaning that it always finds an optimal solution if one exists. Second, A* can take advantage of an \emph{admissible heuristic}, $h(v)$,  to find the optimal path faster. An admissible $h(v)$ provides an estimate of the cost from a vertex $v$ to the optimal goal vertex that is always less or equal to the actual cost, i.e., it must never overestimate the cost. For any given admissible heuristic, A* is {optimally efficient}, i.e., no other complete search algorithm could search fewer vertices with the same heuristic.

The heuristic function is problem specific: in our system, the cost of a path to $v$ is the cost of the schedule in $v_s$, $cost(R,v_s)$, thus it  is calculated differently for different performance goals $R$. Hence satisfying the admissiblity requirement depends on the semantics of the performance metric. Here, we distinguish \emph{monotonically increasing} performance metrics from those that are not. A performance goal is monotonically increasing if and only if the penalty incurred by a schedule $u_s$ never decreases when adding queries. Formally, at any assignment edge connecting $u$ to $v$, $\Penalty{R}{v_s} \geq \Penalty{R}{u_s}$. Maximum query latency is monotonically increasing performance metric, since adding an additional query on the queue of the last provisioned VM will never decrease the penalty. Average latency is not monotonically increasing, as adding a short  query may decrease the average latency and thus the penalty. For monotonically increasing performance goals, we define a heuristic function $h(v)$ that calculates the cheapest possible runtime for the queries that are still unassigned at $v$. In other words, $h(v)$ sums up the cost of the cheapest way to process each remaining query by assuming VMs could be created for free\footnote{\scriptsize For performance goals that are not monotonically increasing, we do not use a heuristic, which is equivalent to using the null heuristic, $h(v) = 0$.}:
\begin{equation}
\label{eq:h}
h(v) = \sum_{q^x_y \in v_u} \min_i \left[ f^i_r \times l(q^x_y, i) \right]
\end{equation}
\begin{lemma}
For monotonically increasing performance goals, the search heuristic defined in Equation~\ref{eq:h} is admissible. 
\end{lemma}
\begin{proof}
Regardless of performance goals, query classes, or VM performance, one always has to pay the cost of renting a VM for the duration of the queries. More formally, every path from an arbitrary vertex $v$ to a goal vertex must include an assignment edge, with cost given by Equation~\ref{eq:weight}, for each query in $v_u$ and for some VM type $i$. When the performance goal is monotonically increasing, the term $\Penalty{R}{v_s} - \Penalty{R}{u_s}$ is never negative, so $h(v)$ is never larger than the actual cost to the goal vertex.
\end{proof}

\subsection{Feature Extraction}
After we have generated the optimal schedules for each of the sampled workloads, we generate the training set for our decision tree classifier. The training set consists of $(decision, features)$ pairs indicating the decisions that was made by A* while calculating optimal schedules and performance/workload related features at the time of the decision. Each decision represents an edge in the search graph, and is therefore a decision to either (a) create a new VM of type $i$, or (b) assign a query of template $T_j$ to the most recently created VM. Thus, the domain of possible decisions is equal to the sum of the number of query templates and the number of VM types.  

We map each decision (edge) $(u,v)$ in the optimal path to a set of features of its origin vertex $u$ since there is a correspondence between a vertex and the optimal decision made at that vertex.  Specifically, for a given vertex $u$, the edge selected in the optimal path is independent of $u$'s parents or children but depends on the unassigned queries $u_u$ and the schedule so far, $u_s$. However, the domains of $v_u$ and $v_s$ are far too large to enumerate and do not lend themselves to machine learning algorithms ($v_u$ and $v_s$ are neither numeric or categorical). Therefore, we extract features from \emph{each} of the vertices in \emph{all}  the optimal paths we collected for \emph{all} of the $N$ sample workloads.

{\bf Feature Selection} One of the main challenges of our framework is to identify a set of efficiently-extracted features that can lead to a highly effective decision model. Here, the space of candidate features is extremely large, as one could  extract features relating to queries (e.g., latency, tables, join, predicates, etc.), the underlying VM resources (CPU, disk, etc.), or combinations thereof. Since even enumerating all possible features would be computationally difficult, finding the optimal set of features is not feasible. Therefore, we focus on features that (1) are fast to extract, (2) capture performance and cost factors that may affect the scheduling decisions, and (3) have appeared in existing workload scheduling and resource provisioning heuristics~\cite{icbs,pmax,sci_place}.

Surprisingly, many ``common sense'' features did not prove to be effective. For example, we extracted the number of available queries of a certain template $X$ still unassigned, \texttt{unassigned-X}. Since each training sample is small, the training set only contained small values of \texttt{unassigned-X}, so the decision tree did not learn how to handle the very large values of \texttt{unassigned-X} encountered at runtime. We find that effective features must be \emph{unrelated to the size of the workload}, since the large workload sizes encountered at runtime will not be represented in the training set. Additionally, we extracted the number of assigned queries of a certain template $X$, \texttt{assigned-X}. With tight performance goals, this feature is well-represented in the training data because machines ``fill up'' after a small number of queries. However, this feature shares information with \texttt{wait-time}, as the  two are highly related. Because of this redundancy, adding \texttt{assigned-X} to the model did not improve performance. Therefore, effective features must be reasonably \emph{independent from one another} to avoid representing redundant data.

We have experimentally studied numerous features which  helped us form a set of requirements for the final feature set.  First, our selected features should be {agnostic to the specifics of the query templates and performance goals}. This will allow our framework to be customizable and enable it to learn good strategies independently of the query templates and performance metric expressed by the application. 

Second, effective features must be {unrelated to the size of the workload}, since the large workload sizes encountered at runtime will not be represented in the training sample workloads, whose size is restricted in order to obtain the optimal schedules in a timely manner (e.g. tracking the exact number of unassigned instances of a particular template will not be useful, as this value will be very small during training and very large at runtime). 

Third, features must be {independent from one another} to avoid extracting redundant information. For example, the wait time in a given VM is related to the number of queries assigned to that VM, and we observed that including only one of these metrics in our feature list was sufficient to give us low-cost schedules.  

From an information-theoretic point of view, the minimum number of bits needed for an optimal feature set is $\log_2(|T| + |V|)$, where $|T| + |V|$ is sum of the number of query templates and the number of VM types, because this is the minimum number of bits needed to differentiate each edge in the graph. However, extracting such an optimal set of features is not likely to be computationally-efficient because of the complexity of the problem. Since we require feature extraction to be fast, we do not attempt to find a feature set with such low entropy. Computationally-efficient approaches must therefore make use of higher-entropy, easy-to-extract features as guides towards good approximate solutions.

Based on these observations, we extract the following features for  each vertex $v$ along the optimal path:
\begin{CompactEnumerate}
\item{\texttt{wait-time}: the amount of time that a query would have to wait before being processed if it were placed on the most recently created VM. {Formally, \texttt{wait-time} is equal to the execution time of all the queries already assigned to the last VM. \cut{ $\sum_{q^x_y \in head(v_s)} l(q^x_y, i)$ where $i$ is the type of the most recently provisioned VM, $head(v_s)$.} This feature can help our model decide which queries should be placed on the last added VM based on their deadline. For example, if a machine's wait time has exceeded the deadline of a query template, it is likely that no more queries of this template should be assigned to that VM. Alternatively, if a machine's wait time is very high, it is likely that only short queries should be assigned.}} 

\item{\texttt{proportion-of-X}: the proportion of queries on the most recently created VM that are of query template \texttt{X}. {In other words, \texttt{proportion-of-X} is the ratio between the number queries of template \texttt{X} assigned to the VM and the total number of queries assigned to the VM. For example, if the VM currently has four queries assigned to it, with one of those queries being of template $T_1$ and three being of template $T_2$, then we extract the features \texttt{proportion-of-T1=0.25} and \texttt{proportion-of-T2=0.75}. We only need to consider the most recently created VM because the assignment edges in the reduced graph only assign queries to the most recent VM. Since each sample workload contains only a limited number of queries, keeping track of the exact number of instances of each query template would not scale to large workloads. Therefore, we track the proportion instead.}}

\item{\texttt{supports-X}: whether or not the most recently created VM is capable of processing a query of class \texttt{X}. This feature is needed when not all VM types are capable of handling every type of query.} 
\item{\texttt{cost-of-X}: the cost incurred (including any penalties) by placing a query of template \texttt{X} on the most recently created VM, or infinity if the most recently created VM does not support queries of class \texttt{X}.
{\texttt{cost-of-X} is equal to the weight of the outgoing assignment edge for template \texttt{X}. This allows our model to check the cost of placing an instance of a certain query template and, based on its value, decide whether to assign another query to the last rented VM or create a new VM.}}
\item{\texttt{have-X}: whether or not a query of template \texttt{X} is still unassigned.  This feature helps our model understand how the templates of the unassigned queries affects the decisions on the optimal path. If there is no instance of some query template $T_j$ unassigned, then the model places one of the remaining templates. If an instance of $T_j$ exists, the model might prefer to schedule that as opposed to any other template.}
\end{CompactEnumerate}

\XCloud learns models that combine these features into custom-tailored heuristics for specific workloads and performance goals. 

We note that while these features are not enough to \emph{uniquely} identify a vertex and thus learn the \emph{exact} conditions that lead to the optimal schedule, they can shed light on the workload/performance conditions related to the optimal decision.  Furthermore, although we cannot claim that these features will \emph{always} allow \XCloud to learn effective heuristics, our experimental results indicate that these features allow \XCloud to learn a reasonable cross-section of the scheduling algorithm space, and that they are 
expressive enough to generate scheduling strategies capable of efficiently handling commonly used performance goals. One can invent an adversarial performance goal (for example, penalizing all VMs without a prime number of assigned queries) that will fall outside the scope of this feature set.

\subsection{Workload Management Model}
\label{sec:model}

\begin{figure}
	\centering
	\includegraphics[width=0.45\textwidth]{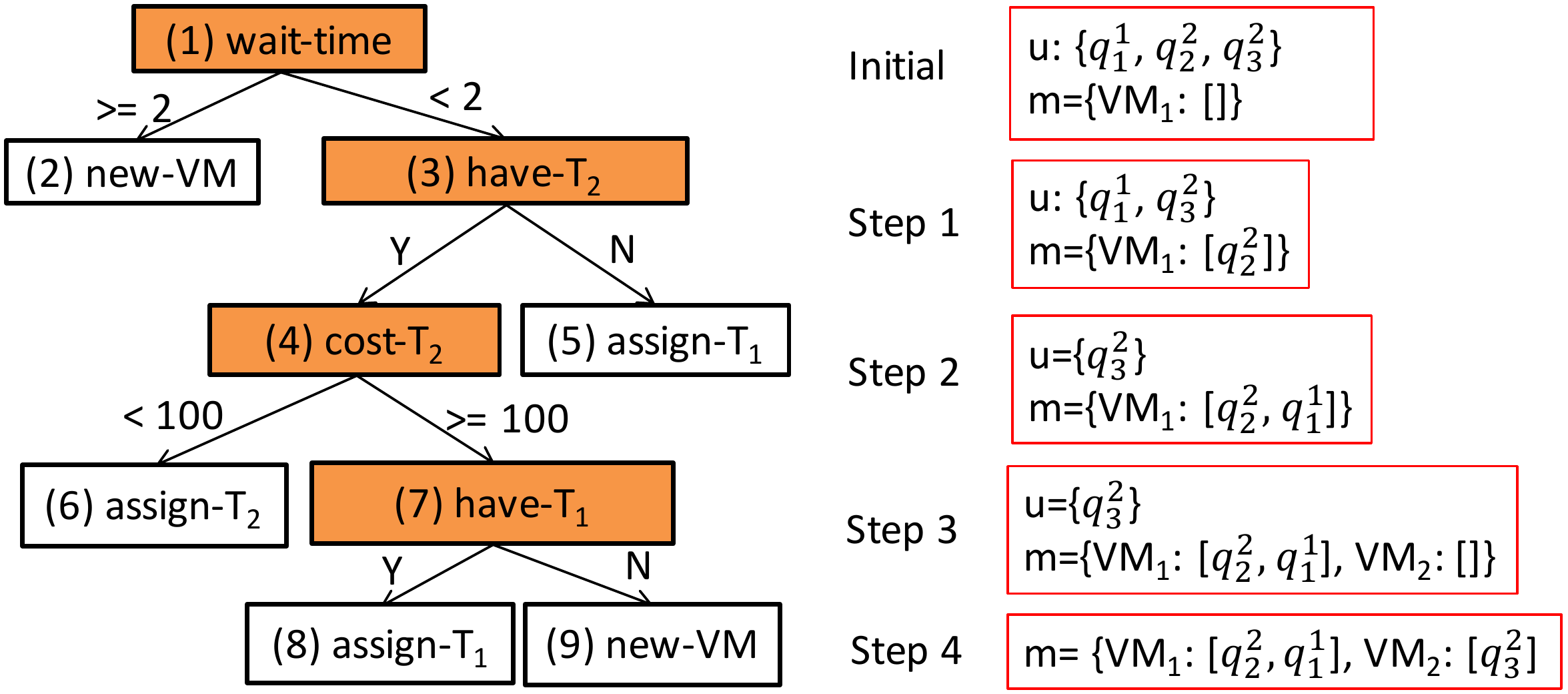}
	\vspace{-1mm}
	\caption{\small An example decision model}
	\vspace{-4.95mm}
	\label{fig:dt}
\end{figure}

Given a training set, WiSeDB  uses a decision tree learner to generate a workload management model. Figure~\ref{fig:dt} shows an example model defined for two query templates $T_1$ and $T_2$.  Each feature node (orange node) of the decision tree represents either a binary split on a numeric or boolean feature. The decision nodes (white nodes) represent the suggested actions.

The decision tree maps a vertex $v$ to an action $a$ by extracting features from $v$ and descending through the tree. Since each action $a$ is represented by a single out-edge of $v$, one can navigate the decision tree by traversing the edge corresponding to the action selected.

 The right side of Figure~\ref{fig:dt} shows how the decision tree is used to come up with the schedule for a workload $Q=\{q^1_1,q^2_2,q^2_3\}$. Each query in $T_1$ has a latency of two minutes and the goal is to execute it within three minutes. Instances of $T_2$  have a latency of one minute and the goal is for each instance to be completed within one minute. For simplicity, we assume  VMs of a single type and that queries are executed in isolation.

Given this workload, the tree is parsed as follows. In the first node (1), we check the wait time, which is zero since all queries are unassigned, hence we proceed to node (3). The workload has queries of template $T_2$ and therefore we proceed to node (4). Here we calculate the  cost of placing an instance of $T_2$. Let us assume the cost is is less than 100 (no penalty is incurred), which leads to node (6) which assigns an instance of $T_2$ to the first VM. Since we have more queries in the workload we next re-parse the decision tree. In node (1) we check the wait time on the most recent VM which is now 1 minute (the runtime of queries of $T_2$) so we move to node (3). Since we have one more query of $T_2$ unassigned, we move to (4). Now the cost of assigning a query of $T_2$ is more than 100 since the new query would need to wait for $q^2_1$ to complete (and thus incur a penalty). Hence, we move to node (7) and we check if there are any  unassigned instances of $T_1$. Since there are ($q^1_1$), we assign $q^1_1$ to the last VM. We re-parse the tree in the same way and {by following nodes (1)$\rightarrow$(2)}, then again as (1)$\rightarrow$(3)$\rightarrow$(4)$\rightarrow$(7)$\rightarrow$(9), so we assign the remaining query $q^2_3$ onto a new VM. 

Each model represents a workload scheduling strategy. Given a batch of queries, the model in  Figure~\ref{fig:dt}  will place an instance of $T_2$, then an instance of $T_1$, and then create a new VM. This process will repeat itself until queries of $T_1$ or $T_2$ are depleted from the incoming batch. When all instances from one of the templates are assigned, single instances of the remaining template will be placed on new VMs until none remain. This strategy  is equivalent to first-fit increasing, which sorts queries in decreasing order of latency and places each query on the first VM where the query ``fits'' (i.e., incurs no penalty).

%% file: adaptive.tex
\section{Adaptive Modeling}\label{sec:shift}

\begin{figure}
\centering
\includegraphics[width=0.45\textwidth]{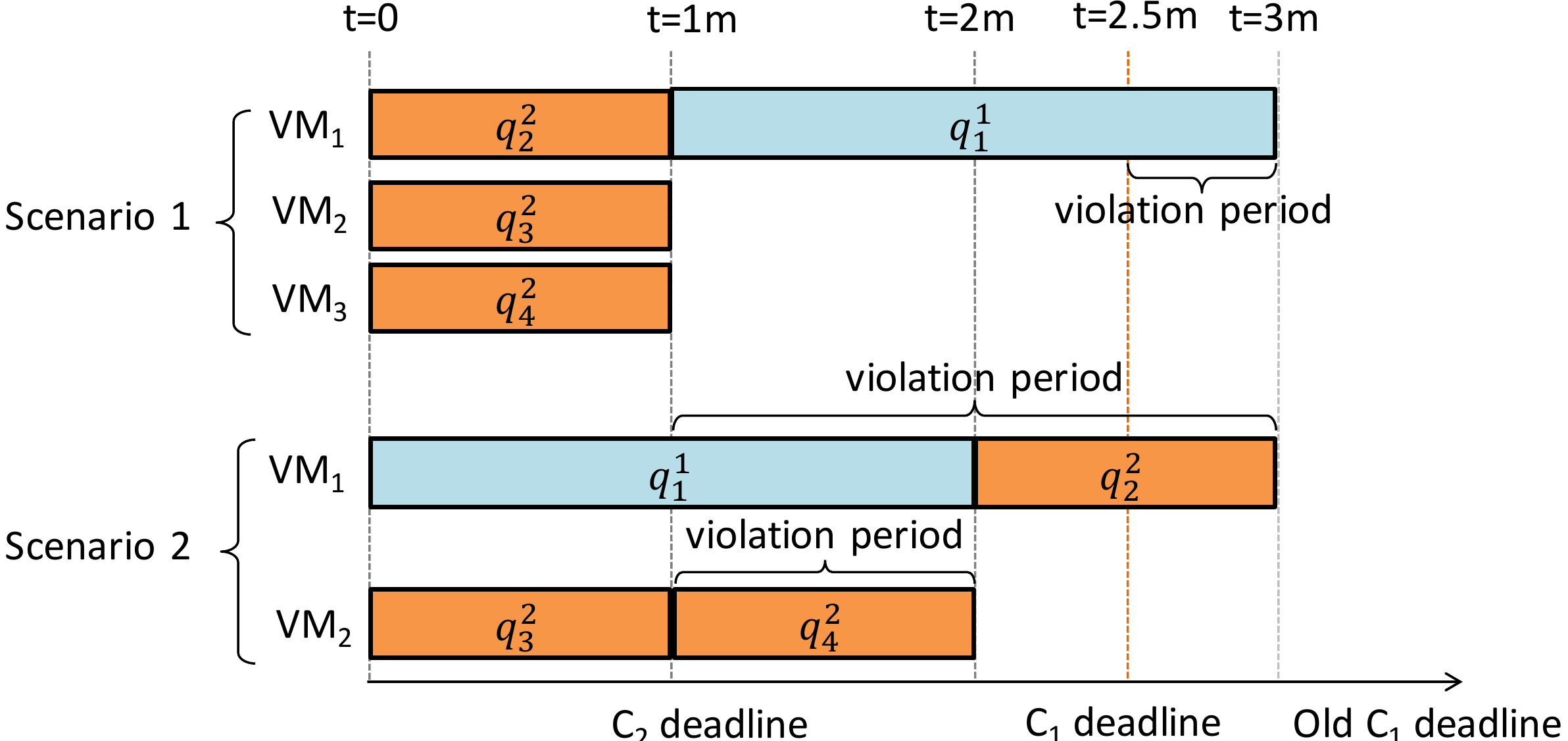}
\caption{{Two different schedules with shifted SLA}}
\label{fig:example_sla_shift}
\end{figure}

It is often desirable to allow users to explore performance/cost trade-offs within the space of possible performance goals~\cite{pslas}. This can be achieved by generating different models for the same workload with stricter/looser performance goals and thus higher/lower costs. However, \XCloud tunes its decision model for a specific goal. Changes in this goal will trigger \XCloud to re-train the model, since changes in performance goal lead to different optimal schedules and hence different training sets. Therefore, generating a set of alternative decision models for numerous performance constraints could impose significant training overhead.

To address this challenge, \XCloud employs a technique that adapts an existing model trained for a given workload and performance goal to a new model for the same workload and stricter performance goals. Our \emph{adaptive modeling} requires little re-training since it leverages the fact that two decision models will share significant parts of their training sets if they were trained for the same query templates $T$ and similar performance goals. Our approach relies on the adaptive A* algorithm~\cite{adaptiveastar}, which reuses information from one graph to create a new search heuristic, $h^\prime(v)$, to search  another graph with identical structure but increased edge weights. 



{\bf Graph Reuse} Let us assume a model $\alpha$ trained for templates $T$ and goal $R$ and a desired new model $\alpha^\prime$  trained for the same  templates $T$ but a different {performance goal} $R^\prime$. 
Without loss of generality, we only consider cases where the performance goal $R^\prime$ is stricter than $R$, since one can start with a substantially loose performance goal than the one requested and restrict it incrementally. 

For example, Figure~\ref{fig:example_sla_shift} shows a modified version of the SLA shown in Figure~\ref{fig:example_sla}. In Figure~\ref{fig:example_sla_shift}, the deadline for query class $C_2$ has been moved up by half a minute, causing Schedule 1 to incur a penalty it did not previously incur.

In order to generate a new model $\alpha^\prime$, WiSeDB re-uses the training data of the existing model $\alpha$ as follows. For each sample workload used for $\alpha$, it modifies its corresponding scheduling graph $G$ by updating the weights to reflect the new performance goal $R'$. Specifically, start-up edges maintain the same cost (the cost to start a new VM), while the cost of placement edges increase due to a stricter performance goals and hence higher penalty fees. Formally, the new cost of an assignment edge $(u, v, q^x_y)$ that places an unassigned query $q^x_y \in u_u$ into the queue of a rented VM in $v_s$ is: \[w(u,v) + \left[\Penalty{R'}{v_s} - \Penalty{R}{v_s}\right] - \left[\Penalty{R'}{u_s}-\Penalty{R}{u_s}\right]\] 

To identify the minimum cost path on the updated graph, we use a new heuristic. For metrics that are monotonically increasing, the new heuristic $h^\prime$ is expressed in terms of the original one $h$ as:
\begin{equation*}
h^\prime(v) = \max \left[h(v), \Cost{R}{g} - \Cost{R}{v}\right]
\end{equation*}
For {performance goals} that are not monotonically increasing, {like average latency,} we simply drop the $h(v)$ term, giving $h^\prime(v) = \Cost{R}{g} - \Cost{R}{v}$.

We use $h^\prime$ to find the optimal schedule for each of $\alpha$'s sample workloads (i.e., we search for the minimum cost paths on each sample's scheduling graph $G$ with the updated weights).  These new paths will serve as training samples for $\alpha^\prime$. \cut{This new training set is then used to learn a decision tree model for the stricter performance goal $R^\prime$.}  
{Intuitively, $h^\prime(v)$ gives the cost of getting from vertex $v$ to the optimal goal vertex under the old performance goal $R$. Since the new performance goal is strictly tighter than $R$, $h^\prime(v)$ cannot overestimate the cost, making it an admissible heuristic. Next, we prove this formally.}

\begin{lemma}
$h^\prime(v)$ is an admissible heuristic, i.e, it does not overestimate the cost of the schedule $v_s$ at the vertex $v$.
\end{lemma}

\begin{proof}

Based on our cost formula (Equation~\ref{eq:cost}), for any schedule $s$ generated by the model $\alpha$, the new performance goal $R'$ will only affect the penalty (i.e., since stricter performance goals on the same schedule can lead only to more violations). Therefore:
\begin{equation}
\label{eq:stricter}
\forall s (\Penalty{R^\prime}{s} \geq \Penalty{R}{s})
\end{equation}
To map this into our scheduling graph, let us consider the set of vertices along an optimal path to a goal vertex $g$ for one sample workload discovered used for the training of $\alpha$. For any particular vertex $v$, we know from Equations ~\ref{eq:cost} and~\ref{eq:stricter} that $\Cost{R^\prime}{v} \geq \Cost{R}{v}$. In other words, the cost of that vertex's partial schedule will cost more if we have a stricter performance goal.  Since this holds for all vertices on the optimal path, it also holds for the optimal goal vertex $g$ under $\alpha$. Hence, if the performance goal becomes stricter, the cost of the final schedule can only increase.

This is illustrated in Figure~\ref{fig:example_sla_shift}: the cost of Scenario $1$ increased when the performance goal was shifted.

Furthermore, for any vertex $v$ along the optimal path to $g$, we know that the minimum possible cost to go from $v$ to $g$ is exactly $\Cost{R}{g} - \Cost{R}{v}$. If the edge weight can only increase due to a stricter performance goal, then the cost to go from $v$ to $g$ under $R$ must be less than or equal to the cost to go form $v$ to $g$ under $R^\prime$. Formally, since $\Cost{R^\prime}{v} \geq \Cost{R}{v}$, it follows that $\Cost{R^\prime}{g} - \Cost{R^\prime}{v} \geq \Cost{R}{g} - \Cost{R}{v}$. 

For example, in the graph shown in Figure~\ref{fig:graph}, $\Cost{R}{B} = 6$ and $\Cost{R}{G} = 26$. The cost to get from $B$ to $G$ is thus $20$. If $R^\prime$ resulted in the assignment $\overline{EG}$ having a higher cost of $100$ (because $q^2_3$ misses its deadline), then it holds that the cost of going from $B$ to $G$ under $R^\prime$ (110) is greater than the cost of going from $B$ to $G$ under $R$: $\Cost{R^\prime}{G} - \Cost{R^\prime}{G} \geq \Cost{R}{G} - \Cost{R}{B}$. 

While the optimal goal vertex for a set of queries may be different under $R$ than under $R^\prime$,  {the cost of the optimal vertex $g'$ under $R^\prime$ is no less than the cost of the optimal vertex $g$ under $R$}, because $g_s$ was optimal under $R$. {Intuitively, if there was some vertex $\gamma$ and complete schedule $\gamma_s$ of a workload under $R^\prime$ with a lower cost than optimal schedule $g_s$, then $\gamma_s$ would also have a lower cost than $g_s$ under $R$, which contradicts that $g_s$ is optimal.} Therefore the cost from any vertex $v$ under $R^\prime$ to the unknown optimal goal vertex $g^\prime$ is no less than the cost to get from $v$ to $g$ under $R$:
\begin{equation*}
\Cost{R^\prime}{g^\prime} - \Cost{R^\prime}{v} \geq \Cost{R}{g} - \Cost{R}{v}
\end{equation*}
Hence $\Cost{R}{g} - \Cost{R}{v}$ is admissible since it never overestimates the actual cost to get from a vertex $v$ to the unknown goal vertex $g'$ under the performance $R'$.
\end{proof}

Since $h^\prime(v)$ is admissible, answers produced by using it are guaranteed by the A* algorithm to be correct~\cite{astar}. The improved heuristic is able to find the optimal solution much faster, as we will demonstrate experimentally in Section~\ref{sec:expr}. This approach saves \XCloud from searching the entire scheduling graph for each new model, which is the step with the dominating overhead.

%% file: runtime.tex
\section{Run Time Functionality}\label{s:runtime}

Using our  adaptive modeling approach, WiSeDB recommends to the application a set of decision models (a.k.a. workload management \emph{strategies}) for scheduling incoming queries. During runtime, the user selects a model with a desirable performance vs. cost trade-off. Given a batch query workload, \XCloud uses the selected model to generate the schedule for that workload. The same model can also be used for online scheduling where queries arrive one at a time. The system remains in this mode until the user either wishes to switch strategy or has no more queries to process.


\subsection{Strategy Recommendation}

While generating alternative decision models, our goal is to identify a small set of $k$ models that represent significantly different performance vs. cost trade-offs. To achieve this we first create a sequence of performance goals in increasing order of strictness, $\mathscr{R} = R_1, R_2, \dots, R_n$, {in which the application-defined goal $R$ is the median}.  For each goal $R_i \in \mathscr{R}$, we train a decision model (by shifting the original model as described in Section~\ref{sec:shift}) and calculate the average cost of each query template over a large random sample workload $Q$. Then, we compute the pairwise differences between the average cost of queries per template of each performance goal $R_i \in \mathscr{R}$ using Earth Mover's Distance~\cite{emd}. We find the smallest such distance between any pairs of performance goals, $EMD(R_i, R_{i+1})$, and we remove $R_{i+1}$ from the sequence. We repeat this process until only $k$ performance goals remain. 

A similar technique was also used in~\cite{pslas} where it was shown that it produces a desirable distribution of performance goals (or ``service tiers'') that represent performance/cost trade-offs. However, unlike in~\cite{pslas}, we do not need to \emph{explicitly} execute any workloads in order to compute the expected cost of a particular strategy. Instead, for each strategy, \XCloud provides a cost estimation function that takes as a parameter the number of instances per query template. Users can easily experiment with different parameters for incoming workloads and can estimate the expected cost of executing these workloads using each of the proposed strategies.

\subsection{Batch Query Scheduling}
Given one of the recommended strategies and a batch of queries to be executed on the cloud, \XCloud uses the strategy and produces a schedule for this workload. A  detailed example of this process was given in Section~\ref{sec:model}.  The schedule indicates the number and types of VMs needed, the assignment of queries to VMs and the query execution  order in each VM. The application is then responsible for executing these scheduling decisions on the IaaS cloud infrastructure. {In the event that the workload includes a query that cannot be exactly matched to one of the templates given in the workload specification, \XCloud will treat the unknown query as an instance of the query template with the closest predicted latency.} While queries of unseen templates are obviously not ideal, this solution ensures that queries of unseen templates will at least be {placed appropriately based on their latency, since two queries with identical latency are identical to \XCloud.}

\subsection{Online Query Scheduling}
\label{sec:online}

\begin{figure}
\centering
\includegraphics[width=0.45\textwidth]{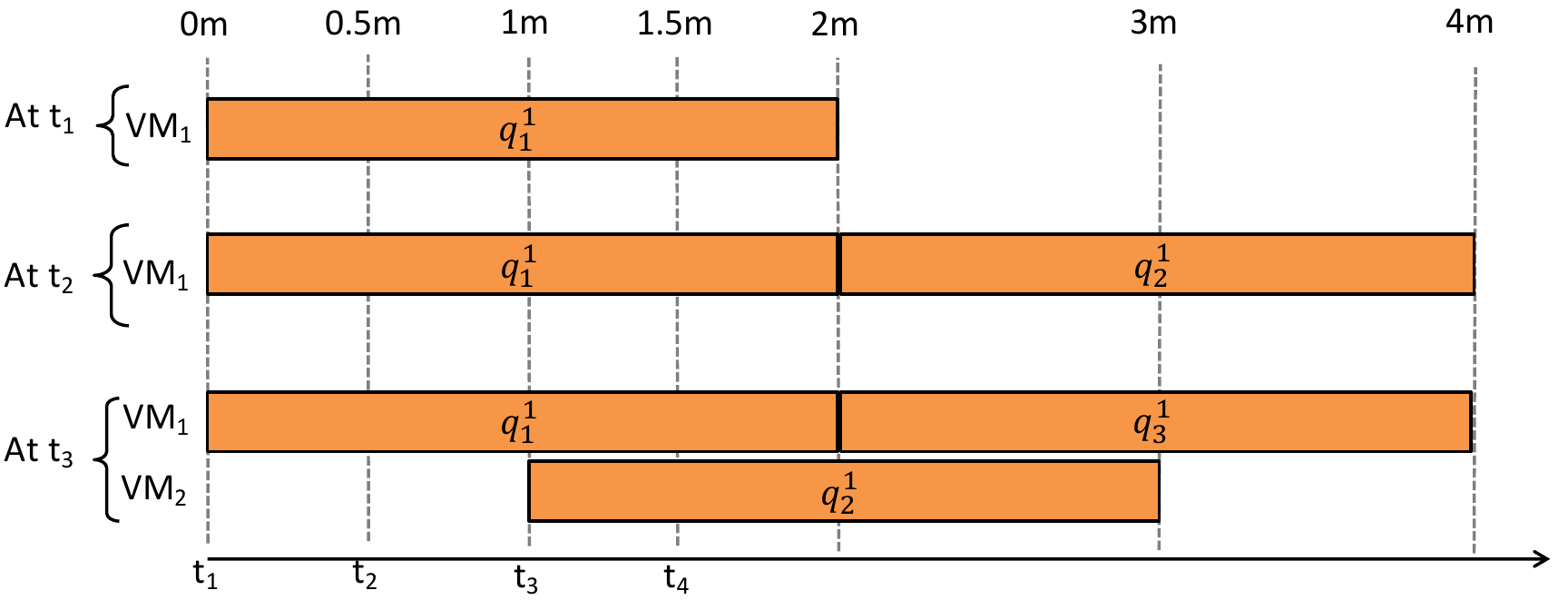}
\vspace{-0.5em}
\caption{\small{Online scheduling example}}
\vspace{-3mm}
\label{fig:example_sla_online}
\end{figure}

\XCloud can also handle non-preemptive online query scheduling. We will describe how \XCloud handles the general case of online scheduling and then we will discuss two optimizations.

{\bf General Online Scheduling} Let us  assume a decision model $\alpha$ trained for a performance goal $R$ and templates $T$. We also assume that a user submits queries, $\{q^x_1, q^y_2, \dots,\}$ at  times $\{t_1, t_2, \dots\}$ respectively. Online scheduling can be viewed as a series of successive batch scheduling tasks where each batch includes a single additional query. The first query is scheduled at time $t_1$ as if it was a batch. When query  $q^x_i$ arrives at $t_i$, we create a new batch $B_i$ containing the queries that have not started executing by $t_i$ and use the model to re-schedule them along with query $q^x_i$.

Scheduling $B_i$ as a batch using the same decision model might not offer low-cost solutions. This is because this batch includes queries that have been sitting in the queue, and hence their expected latency is now higher than what was assumed when training the model. Formally, at any given time $t_i$, any query $q^x_y$ that has yet to run will have a wait time of $(t_i - t_y)$ and thus a  final latency of $l'(q^x_y, k)= l(q^x_y, k) + (t_i - t_y)$, where $k$ is the type of the VM on which $q^x_y$ was assigned and $t_y$ is the arrival time of $q^x_y$. To address this, WiSeDB treats $q^x_y$ as if it were of a ``new'' template which has the same structure at $T_x$ but its expected latency  is $l'(q^x_y, k)$. Then, it trains a new decision model for the augmented template set that includes this ``new'' template. This new model will  provide schedules that account for the elapsed wait time of $q^x_y$.

Figure~\ref{fig:example_sla_online} shows an example. Here, a model $\alpha$ is trained for one template $T_1$. Instances of $T_1$ have a latency of $2m$. \cut{We assume the performance goal is to execute each query within two minutes.} We assume queries arrive in the order: $\{q^1_1, q^1_2, q^1_3\}$. \XCloud first schedules query $q^1_1$, which arrived at $t_1$, as a batch $B_1 =\{q_1^1\}$. At $t_2$, $q^1_2$ arrives and \XCloud creates a batch $B_{2}=\{q^1_2\}$ for it and generates a schedule using model $\alpha$ which assigns it to the first VM, right after the first query. At time $t_3$, when one more query $q^1_3$ arrives, $q^1_2$ has not yet started executing, so  we create a new batch $B_{3}=\{q^1_2,q^1_3\}$. However, even though $q^1_2$ and $q^1_3$ are of the same template, they have different execution times: $q^1_3$ has a latency of $1$ minute, but $q^1_2$ has a latency of $1 + (t_3-t_2)$ minutes since it has been waiting for $(t_3-t_2)$ minutes. Using model $\alpha$, which has been trained to handle queries of with an execution time of one minute, might not produce a desirable (low-cost) schedule for $B_{3}$. So, we train a new model, $\alpha'$, whose workload specification includes an ``extra'' template for instances of $T_1$ with latency of $1 + (t_3-t_2)$ minutes. $\alpha'$ places $q^1_3$ after $q^1_1$ on the first VM and $q^1_2$ onto a new VM.

\subsubsection{Retraining Optimizations}
A substantially fast query arrival rate could rapidly create ``new templates'' as described above, causing a high frequency of retraining. Since training can be expensive, we offer two optimizations to prevent or accelerate retraining.

{\bf Model Reuse} Upon arrival of a new query, \XCloud creates a new decision model which take into account the wait times of the queries submitted so far but not already running. \XCloud strives to reuse models, aiming to reduce the frequency of retraining. In the above example, let us assume a new query $q^1_4$ arrives at $t_4$, at which point we need to generate a schedule for $B_{4} =\{q^1_3, q^1_4\}$ ($q^1_1$ and $q^1_2$ are running). If $(t_3-t_2) = (t_4-t_3)$, then the previously generated model $\alpha^\prime$ can be reused to schedule $B_4$, since the ``new'' template required to schedule $q^1_3$ is the same as the ``new'' template previously used to schedule $q^1_2$.

Next, let us consider the general case. We define $\omega(i)$ to be the difference in time between the arrival of oldest query that has not started processing at time $t_i$ and the arrival of the newest query at time $t_i$. Formally,
\begin{equation*}
\begin{split}
\omega(i) & =  t_i - t_j \mbox{ where } \\
j         & =  \min \{ j \mid j \leq i \land \exists q^x_y (q^x_y \in B_j \land q^x_y \in B_i)\}\\
\end{split}
\end{equation*}

Clearly, the absolute time of a workload's submission is irrelevant to the model: only the difference between the current time and the time a workload was submitted matters. Formally, if $\omega(i) = \omega(j)$, then $B_i$ and $B_j$ can be scheduled using the same model.  {In practice, $B_i$ and $B_j$ can be scheduled using the same model if the difference between $\omega(i)$ and $\omega(j)$ is less than the error in the query latency prediction model.} By keeping a mapping of $\omega(x)$ to decision models\footnote{\scriptsize{Experimentally, we found that this mapping can be stored using  a few MB since each decision tree is relatively small.}}, \XCloud can significantly reduce the amount of training that occurs during online scheduling. 

\cut{For example, for $t_5$ and $t_6$ in Figure~\ref{fig:example_sla_online}, $\omega(5) = 0.5$ and $\omega(6) = 0.5$. Therefore, $B_5$ and $B_6$ can be scheduled using the same model.} 

{\bf Linear Shifting} For some performance metrics, scheduling queries that have been waiting in the queue for a certain amount of time is the same as scheduling with a stricter deadline. This can significantly reduce the training overhead as the new models could reuse the scheduling graph of the previous ones as described in Section~\ref{sec:shift}.  Consider a performance goal that puts a deadline on each query to achieve a latency of three minutes. If a query is scheduled one minute after submission, that query can be scheduled as if it were scheduled immediately with a deadline of two minutes.

This optimization can be applied only to \emph{linearly shiftable} performance metrics. 
In general, we say that a metric $R$ is {linearly shiftable} if the penalty  incurred under $R$ for a schedule which starts queries after a delay of $n$ seconds is the same as the penalty incurred under $R^\prime$ by a schedule that starts queries immediately, and where $R^\prime$ is a tightening of $R$ by some known function of $n$. For the metric of maximum query latency in a workload, this function of $n$ is the identity: the penalty incurred under some goal $R$ after a delay of $n$ seconds is equal to the penalty incurred under $R^\prime$, where $R^\prime$ is $R$ tightened by $n$ seconds. For linear shiftable performance metrics, training a new model when new queries arrive can be reduced to shifting a decision model, as described in Section~\ref{sec:shift}.

%% file: experiments.tex
\section{Experimental Results}\label{sec:expr}
Next we present our experimental results, which focus on evaluating \XCloud's \emph{effectiveness} to learn low-cost workload schedules for a variety of performance goals, as well as its runtime \emph{efficiency}.

\subsection{Experimental Setup}\label{s_setup}

We implemented \XCloud using Java 8 and installed it on an Intel Xeon E5-2430 server with  32GB of RAM. The service generates scheduling strategies for  a database application deployed on 6 Intel Xeon E5-2430 servers that can host up to 24 VMs with 8GB of RAM each. This private cloud emulates Amazon AWS~\cite{url-amazonAWS} by using query latencies and VM start-up times measured on \texttt{t2.medium} EC2 instances. By default, our experiments assume a single type of VM unless otherwise specified. 

Our database application stores a 10GB configuration of the TPC-H~\cite{url-tpch} benchmark on Postgres~\cite{url-postgres}. Query workloads consist of TPC-H templates 1 - 10. These templates have a response time ranging from 2 to 6 minutes, with an average latency of 4 minutes.  We set the query processing cost to be the price of that instance ($f_r = \$0.052$ per hour), and we measured its start-up cost experimentally based on how long it took for the VM to become available via connections after it entered a ``running'' state ($f_s = \$0.0008$). 

We trained our models on $N = 3000$ sample workloads with a size of $m = 18$ queries per workload, with queries executed in isolation. Higher values for $N$ and $m$ added significant training overhead without improving the effectiveness of our models while lower ones resulted in poor decision models.  We generated our models using the {J48~\cite{url-weka}} decision tree algorithm and used them to generate schedules for incoming workloads. Our experiments vary the size of these workloads as well as the distribution of queries across the templates. Each experiment reports the average cost over 5 workloads of a given size and query distribution. 

{\bf Performance Metrics}. Our experiments are based on four  performance goals: (1) {\bf Max} requires the maximum query latency in the workload to be less than $x$ minutes. By default, we set $x=15$,  which is 2.5 times the latency of the longest query in our workload. There is a charge of 1 cent per second for any query whose latency exceeds $x$ minutes. (2) {\bf PerQuery} requires that each query of a given template not exceed its expected latency by a factor of $x$.  By default, we set $x=3$ so that the average of these deadlines is approximately $15$ minutes, which is  2.5 times the latency of the longest query. There is a charge of 1 cent per second in which any query exceeds $x$ times its predicted latency. (3) {\bf Average} requires that the average latency of a workload is $x$ minutes. We set $x=10$ so that this deadline is 2.5 times the average query template latency. There is a charge of a number cents equal to the difference between the average latency of the scheduled queries and $x$. (4) {\bf Percent} requires that $y\%$ of the queries in the workload finish within $x$ minutes. By default, we set $y=90$ and $x=10$. If $y\%$ of queries finish within $x$ minutes, there is no penalty. Otherwise, a penalty of one cent per second is charged for time exceeding $x$ minutes.

\subsection{Effectiveness Results}
\label{s:effective}

To demonstrate the effectiveness and versatility of our approach, we compare schedules generated by \XCloud with optimal schedules and known heuristics that have been shown to offer low-cost solutions for specific performance metrics. 

\begin{figure}
  \centering
  \includegraphics[width=0.45\textwidth]{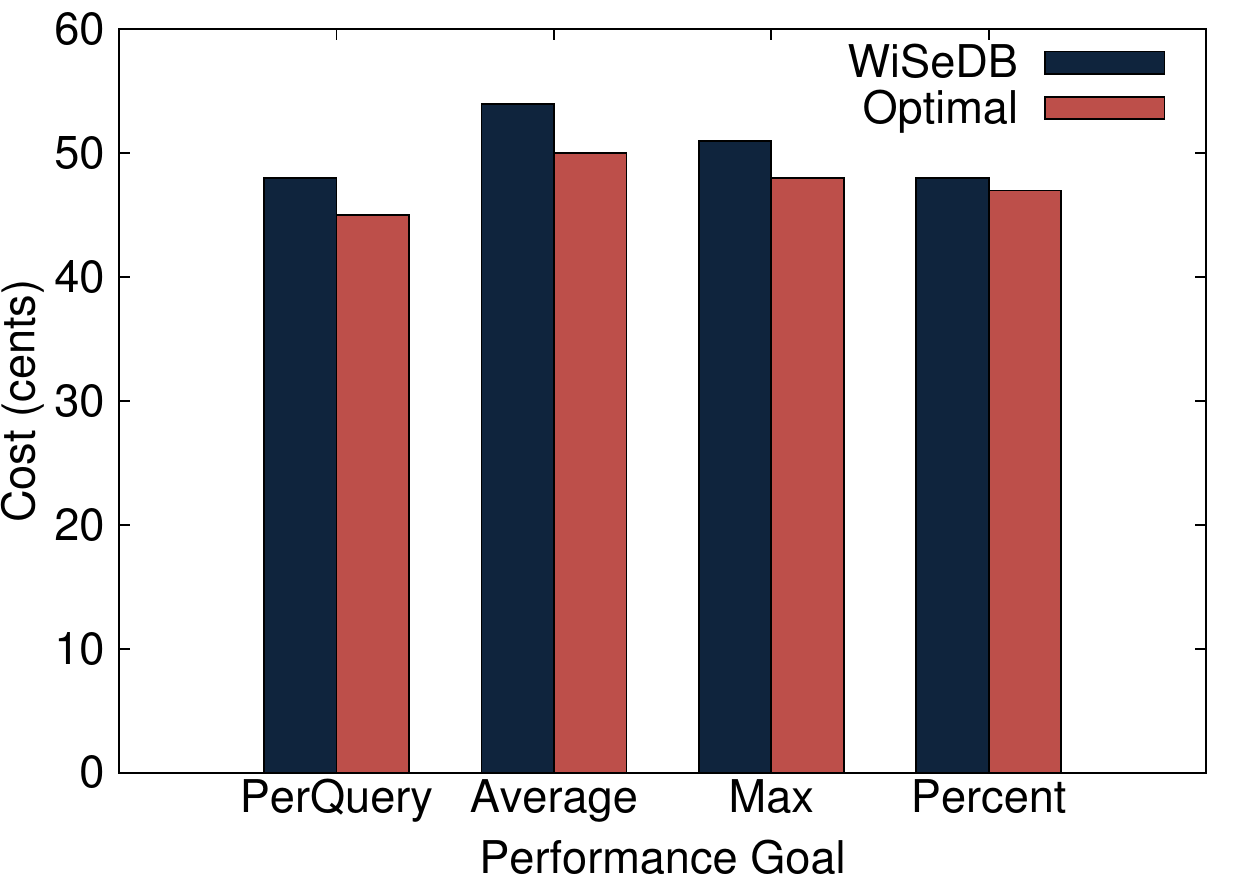}
  \caption{\small Optimality for various performance metrics}
  \label{fig:performance}
\end{figure}

{\bf Optimality} Since our scheduling problem is computationally expensive, calculating an optimal schedule for large workloads is not feasible. However, for smaller workload sizes, we can exhaustively search for an optimal solution. Figure~\ref{fig:performance} shows the final cost for scheduling workloads of 30 queries uniformly distributed across 10 query templates. We compare schedules generated by {\XCloud} with the optimal schedule  ({\tt Optimal}) for all our performance metrics.  {Our  schedules are within $8\%$ of the optimal for all metrics}.

\begin{figure}
  \centering
  \includegraphics[width=0.45\textwidth]{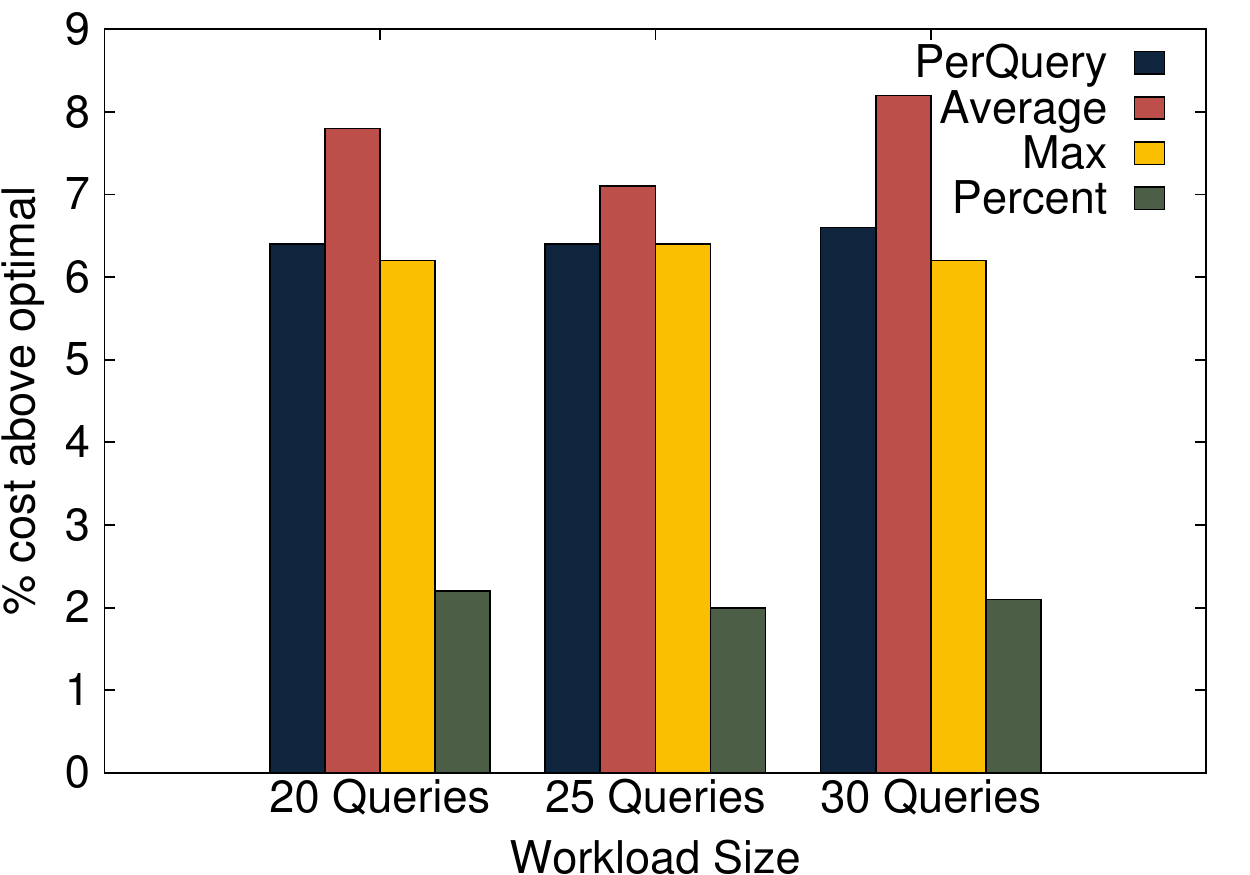}
  \caption{\small Optimality for varying workload sizes}
  \label{fig:performance_scale}
\end{figure}

Figure~\ref{fig:performance_scale} shows the percent-increase in cost over the optimal for three different workload sizes. \XCloud consistently performs within 8\% from the optimal independently of the size of the workload, while for the {\tt Percent} metric, \XCloud learns scheduling heuristics that cost less than 2\% above the best solution.

\begin{figure}
  \centering
  \includegraphics[width=0.45\textwidth]{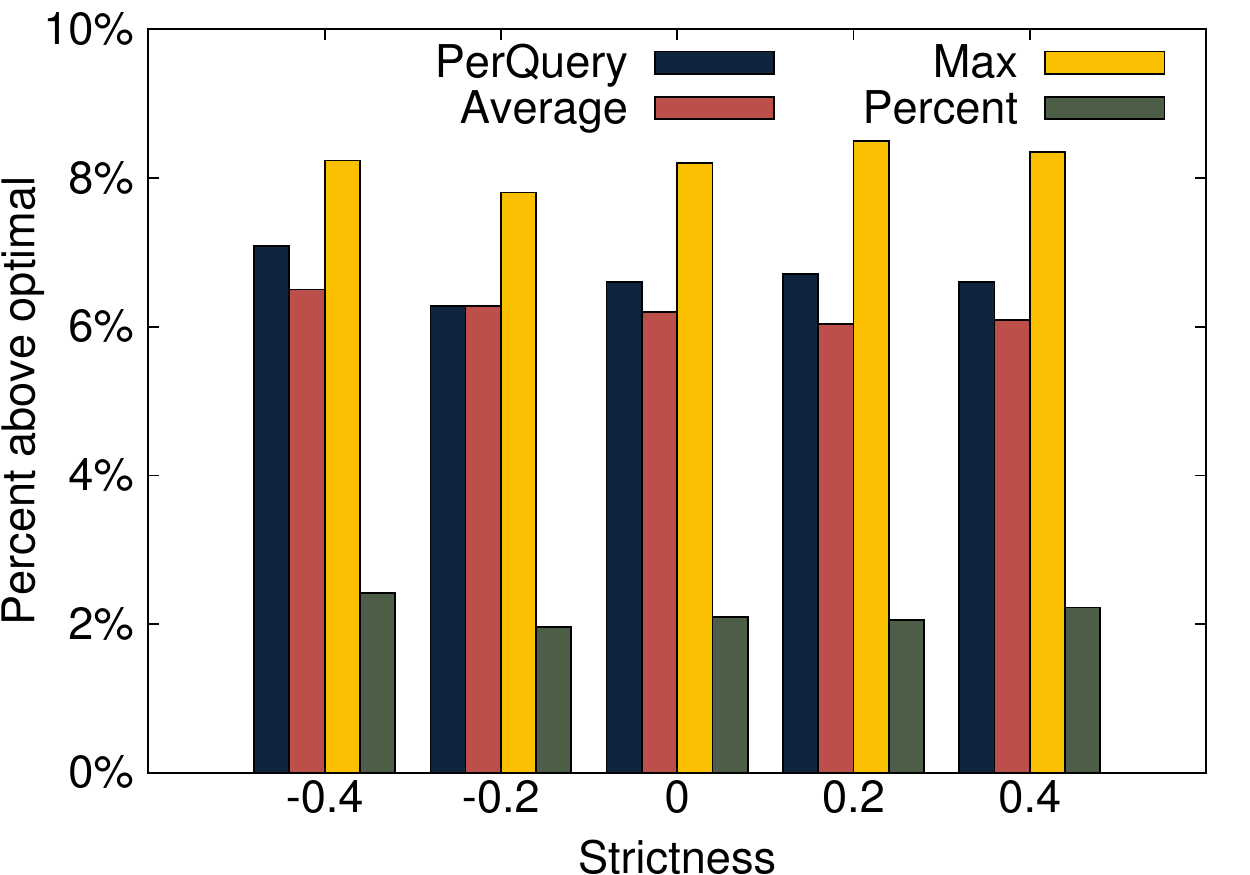}
  \caption{\small {Optimality for varying constraints}}
  \label{fig:strict}
\end{figure}

Figure~\ref{fig:strict} shows the percent-increase in cost over the optimal when we tighten the constraints of the performance goals. To capture this parameter, we introduce a \emph{strictness factor}.  {A strictness factor of} $0$ means a performance goal equal to those described in Section~\ref{s_setup}, whereas a strictness factor $x<0$ represents a performance goal that is only $x$ times as strict as those described above (so it has a more relaxed constraint). A strictness factor of $x>0$ corresponds to a performance goal that is $x$ times stricter than those above (so it has a tighter constraint). Figure~\ref{fig:strict} shows that the strictness factor does not affect the effectiveness of \XCloud relative to the optimal. In the rest of the experiments, we will use performance goals with a strictness factor of $0$.

\emph{We conclude that \XCloud is able to learn effective strategies for scheduling incoming TPC-H workloads for various performance metrics independently of the size of the runtime workloads and the strictness of the performance goal.}

\begin{figure}
  \centering
  \includegraphics[width=0.45\textwidth]{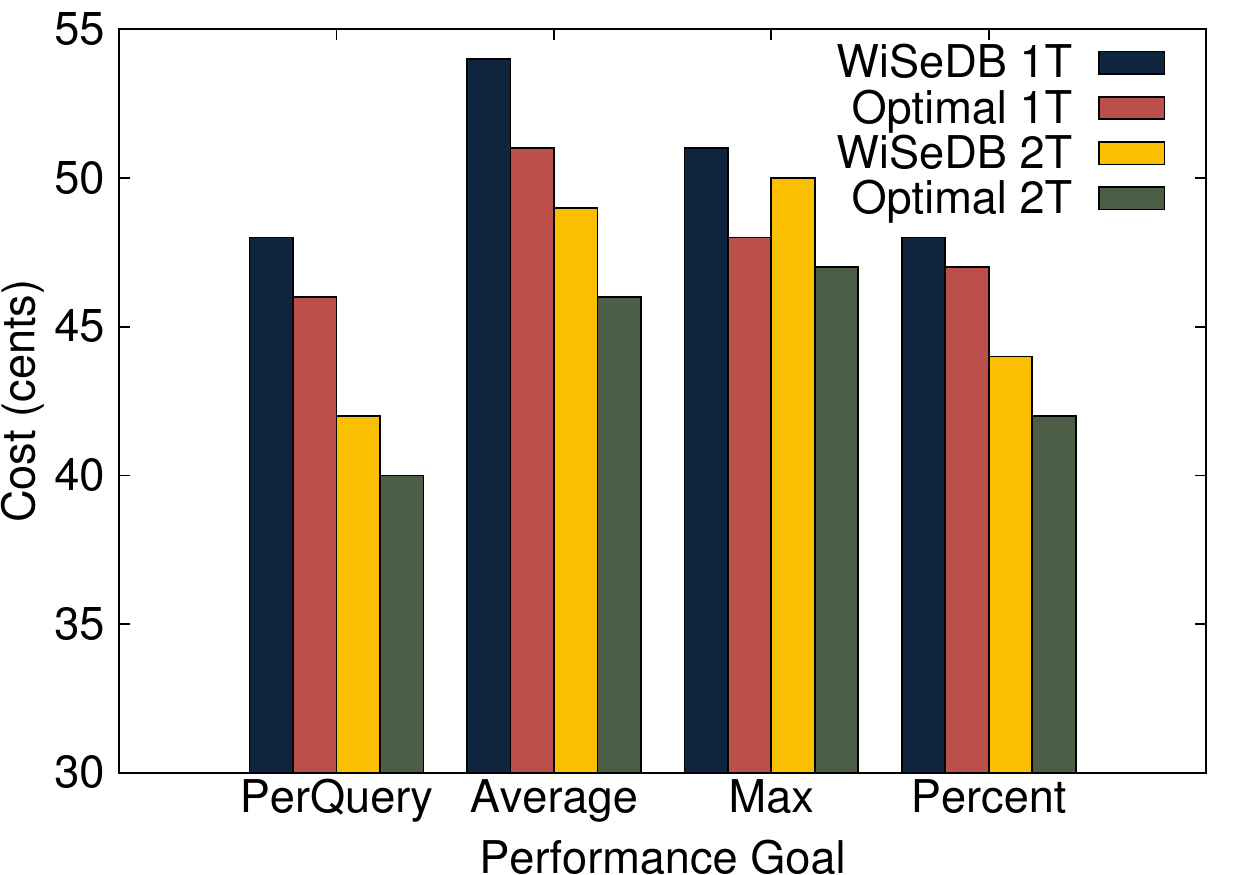}
  \caption{ \small Optimality for multiple VM types}
  \label{fig:multi_vm}
\end{figure}
\begin{figure}
  \centering
  \includegraphics[width=0.45\textwidth]{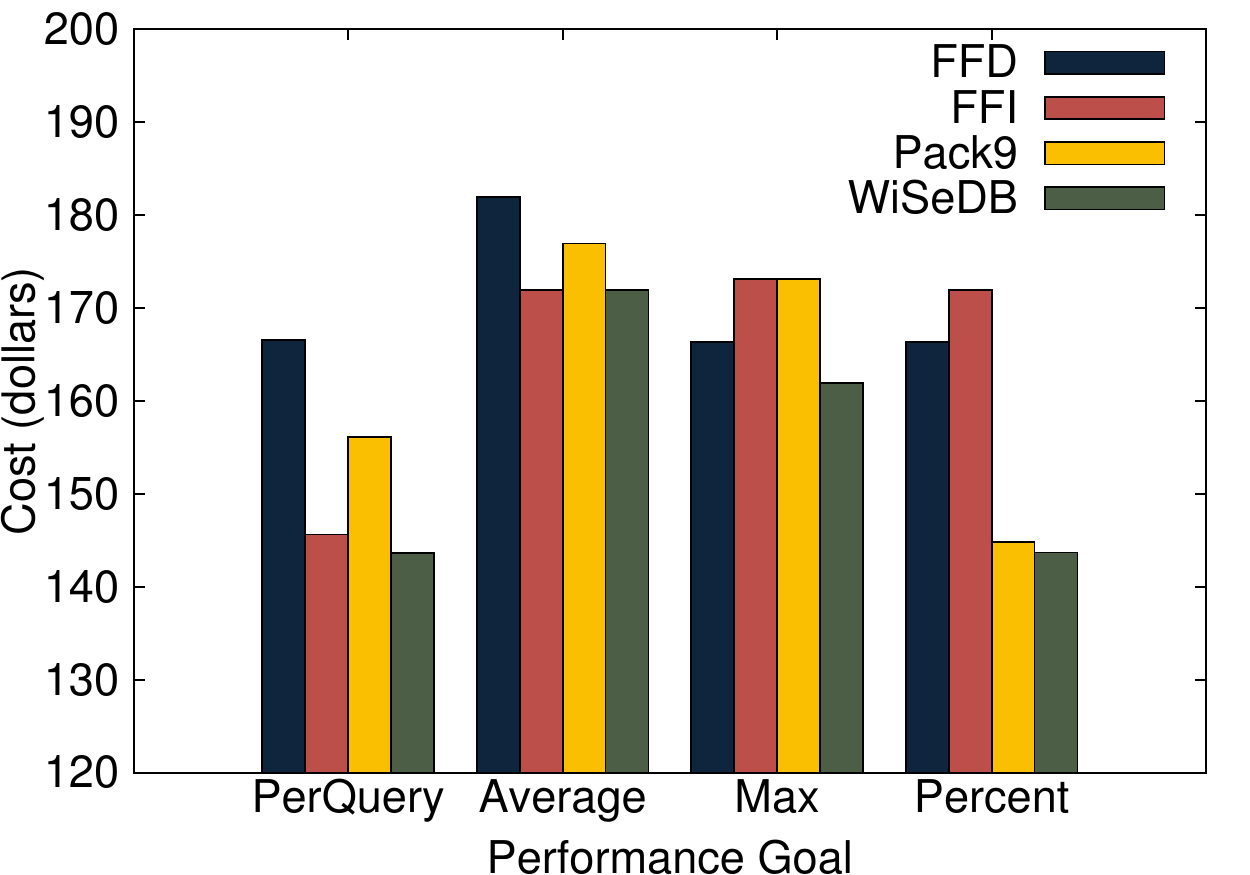}
  \caption{ \small Comparison with metric-specific heuristics}
  \label{fig:performance5000}
\end{figure}

{\bf Multiple VM Types} We also trained decision models assuming the availability of multiple VM types. Figure~\ref{fig:multi_vm} shows the cost of \XCloud schedules against the optimal schedule with one {\tt (\XCloud 1T)} and two {\tt (\XCloud 2T)} VM types. The first type is the \texttt{t2.medium} EC2 type and the second is the \texttt{t2.small} type.  With the TPC-H benchmark, queries that require less RAM tend to have similar performance on \texttt{t2.medium} and \texttt{t2.small} EC2 instances. Since \texttt{t2.small} instances are cheaper, it makes sense to place low-RAM queries on \texttt{t2.small} instances. The results reveal that even when the learning task is more complex (using more VM types adds more edges to the scheduling graph and thus more complexity to the problem), \XCloud is able to learn these correlations and generate schedules that perform within 6\% of the optimal on average. Additionally, the performance always improved when the application was given access to a larger number of VM types. \emph{Hence, \XCloud is able to leverage the availability of various VM types, learn their impact on specific performance goals, and adapt its decision model to produce low-cost schedules}.

{\bf Metric-specific Heuristics} We now evaluate WiSeDB's effectiveness in scheduling large workloads of $5000$ queries. Here, an exhaustive search for the optimal solution is infeasible, so we compare \XCloud's schedules with schedules produced by heuristics designed to offer good solutions for each performance goals. \emph{First-Fit Decreasing} ({\tt FFD}) sorts the queries in descending order and then places each query into the first VM with sufficient space. If no such VMs can be found, a new one is created. FFD is often used as a heuristic for classic bin-packing problems~\cite{ffd}, indicating that it should perform well for the {\tt Max} metric. \emph{First-Fit Increasing} ({\tt FFI}) does the same but first sorts the queries in ascending order, which works well for the {\tt PerQuery}  and the {\tt Average} query latency metrics~\cite{appalg}. {\tt Pack9} first sorts the queries in ascending order, then repeatedly places the 9 shortest remaining queries followed by the largest remaining query. Pack9 should perform well with the {\tt Percent} performance goal because it will put as many of the most expensive queries into the $10\%$ margin allowed.

\begin{figure}
  \centering
  \includegraphics[width=0.45\textwidth]{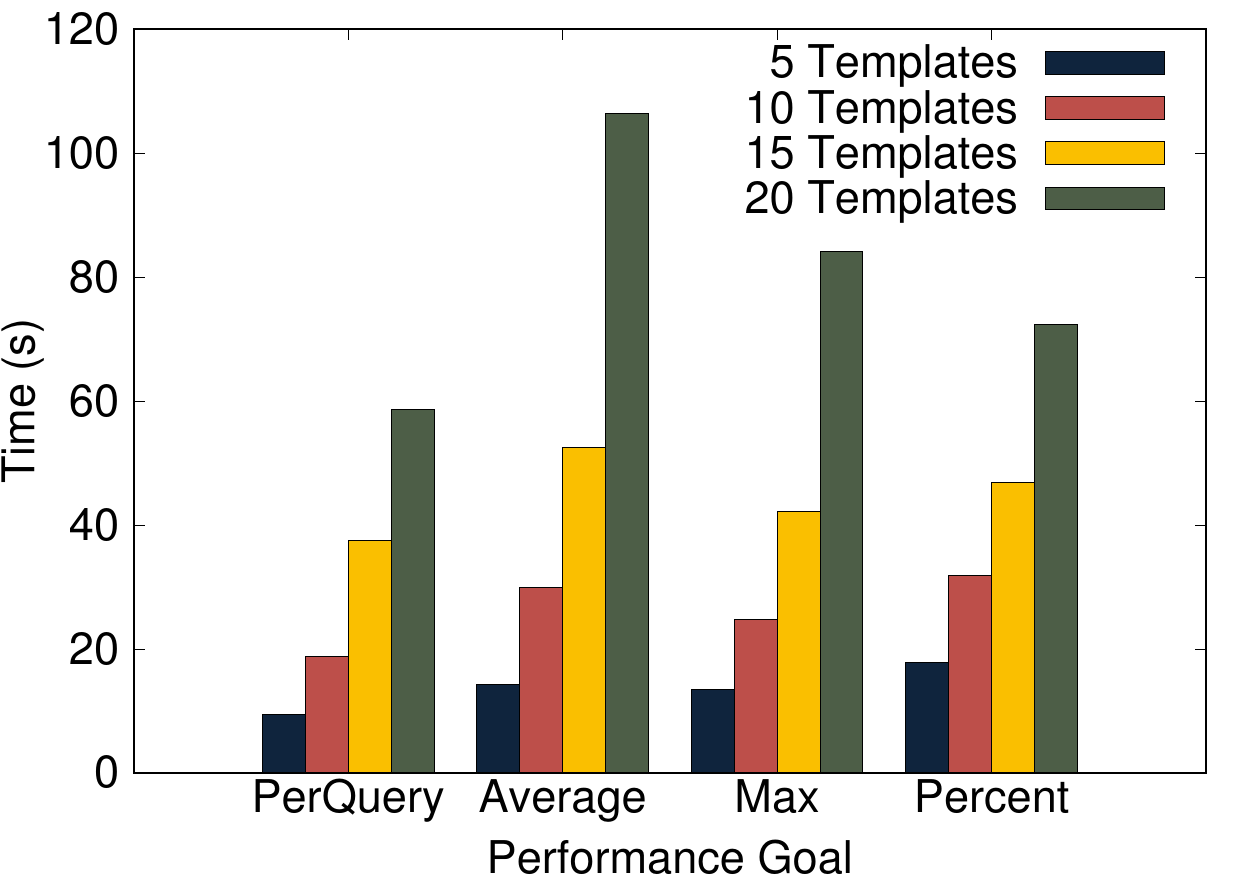}
  \caption{\small Training time vs. \# of query templates}
  \label{fig:training}
\end{figure}
\begin{figure}
  \centering

  \includegraphics[width=0.45\textwidth]{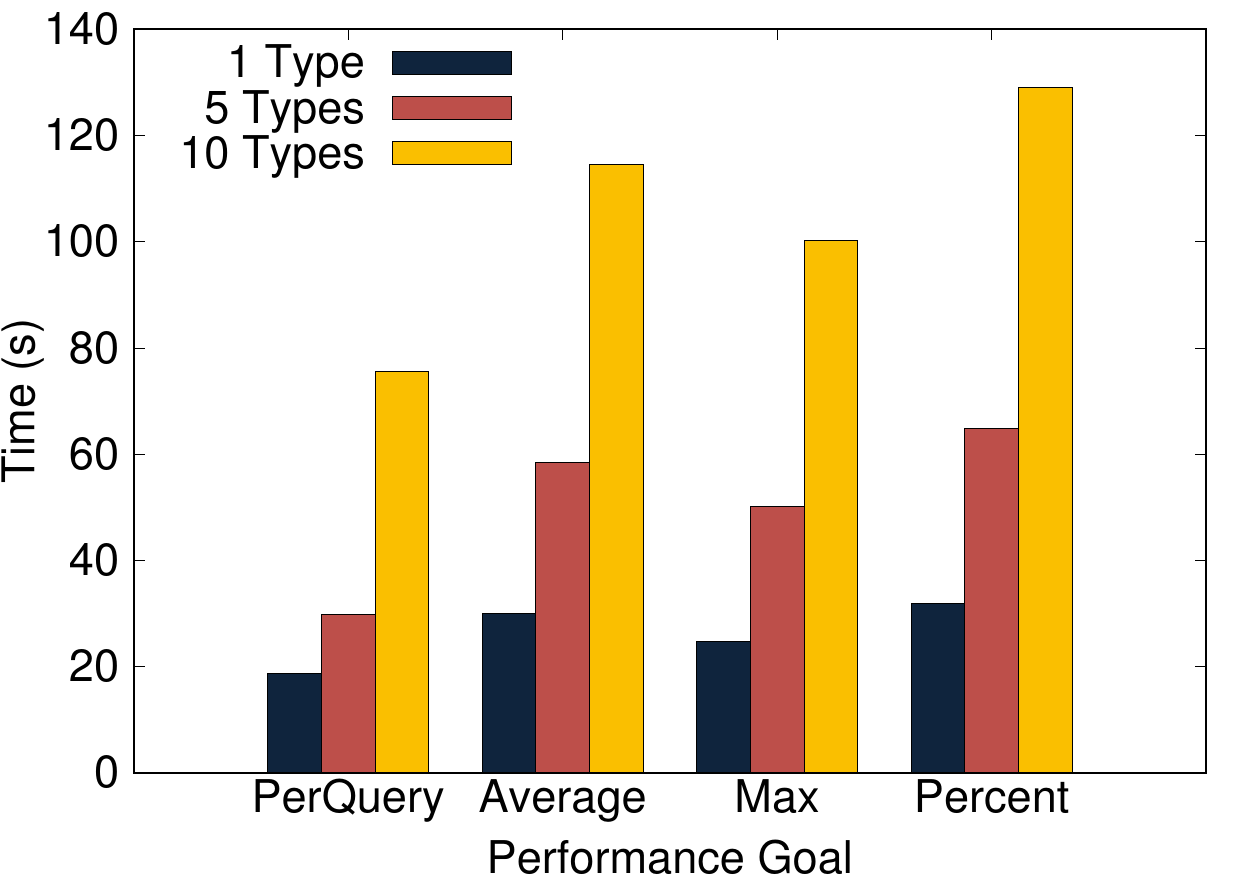}
  \caption{\small Training Time vs. \# of VM types}
  \label{fig:train_vms}
\end{figure}
\begin{figure}
  \centering

  \includegraphics[width=0.45\textwidth]{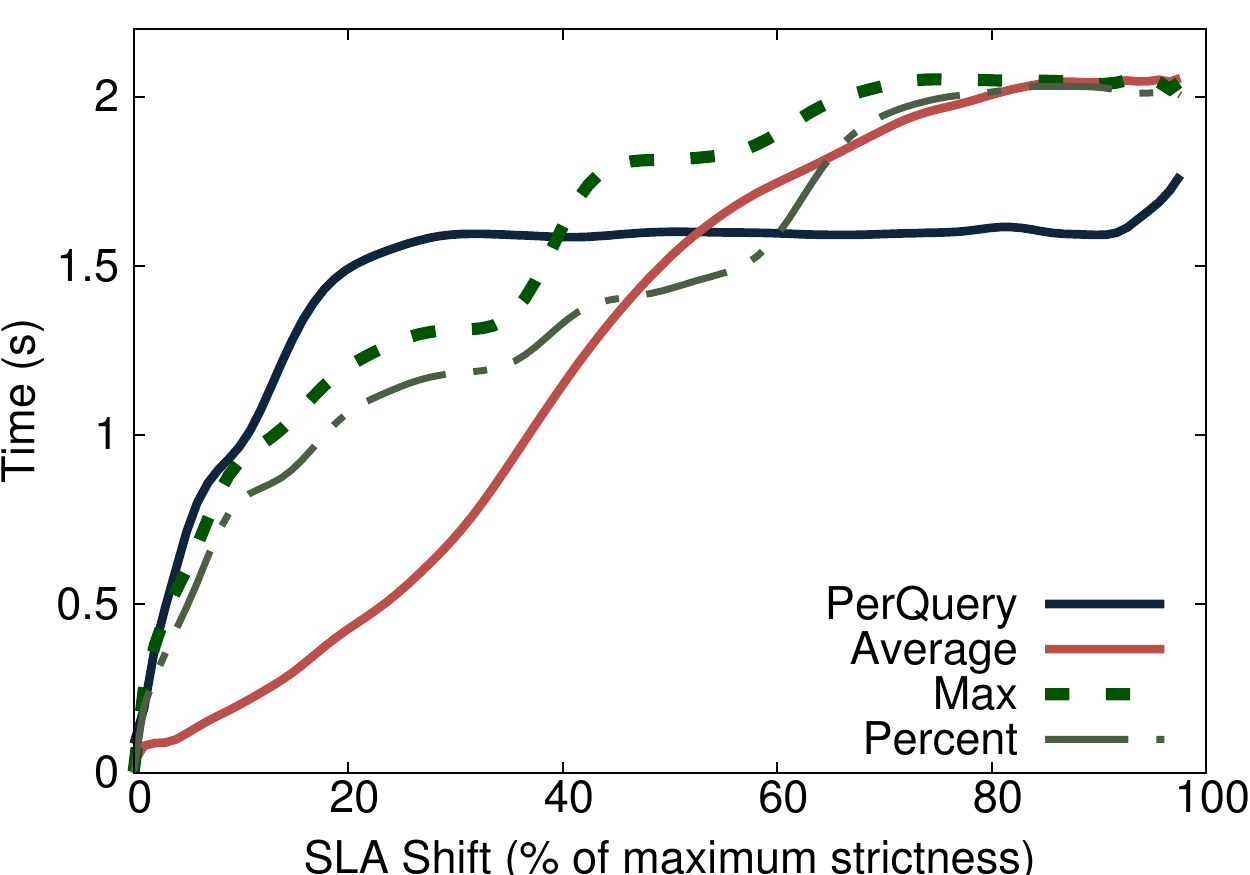}
  \caption{\small Overhead of adaptive modeling}
  \label{fig:retrain}
\end{figure}

 Figure~\ref{fig:performance5000} shows the performance of \XCloud compared to these approaches. The results show that there is no single simple heuristic that is  sufficient to handle diverse performance goals. 
Our service offers schedules that consistently perform better than all other heuristics.  \emph{This indicates that \XCloud outperforms standard metric-specific heuristics, i.e., its training features are effective in characterizing optimal decisions and learning  special cases of query assignments and orderings with the VMs that fit better with each performance goal}. An example of such a special case was given in Section~\ref{sec:model}.

\subsection{Efficiency Results}

{\bf Training Overhead} \XCloud trains its decision models offline. The training time depends on the number of templates in the workload specification as well as the number of different VM types available through the IaaS provider. Figure~\ref{fig:training} shows how long it takes to train models for our four performance metrics, a single VM type,  and varying numbers of query templates. {For this experiment, we used additional query templates from the TPC-H benchmark.} Here, the higher the number of query templates,  the longer the training process since additional query templates represent additional edges that must be explored in the scheduling graph. In the most extreme cases, training can take around two minutes. In more tame cases, training takes less than 20 seconds. {In Figure~\ref{fig:train_vms}, we fix the number of query templates at $10$ and vary the number of VM types available. Again, at the extreme ends we see training times of up to two minutes, with tamer multi-VM type cases taking only $30$ seconds.} \emph{Hence, \XCloud can learn metric-specific  strategies in timely manner while each model needs to be trained once offline and can be applied to any number of incoming workloads}.

\XCloud can adaptively train decision models by tightening the performance goal  of its original model (Section~\ref{sec:shift}). Figure~\ref{fig:retrain} shows the retraining time when tightening the performance constraint by a factor of $p$. In general, we tighten a performance goal by a percentage $p$ using the formula $t + (g - t) * (1-p)$, where $t$ is the strictest possible value for the performance metric, and $g$ is the original constraint (described in Section~\ref{s_setup}). For example, the \texttt{Max} goal has an original deadline of $15$ minutes, and, since the longest template in our workload specification is $6$ minutes, the strictest possible deadline is $6$ minutes. Tightening the \texttt{Max} goal by 33\% means decreasing the deadline from $15$ to $12$ minutes. 

Figure~\ref{fig:retrain} shows that all four metrics can be tightened by up to 40\% in less than a second. Tightening a constraint by a larger percentage takes more time since the number of training samples that have to be retrained increases. This is most immediately clear for the \texttt{Max} metric. The jump at $37\%$ shift represents a large portion of the training set that needs to be recalculated. With tightening by only $20\%$, the optimal schedules used for training do not need to be modified, but a tightening by $40\%$ causes more violations and hence new optimal schedules need to be re-calculated.  \texttt{PerQuery}  and \texttt{Percent} have  curves that behave similarly.

The \texttt{Average} metric is slightly different. Since the query workloads are drawn uniformly from the set of query classes, their sum is normally distributed (central limit theorem). The average latency of each training sample is thus also normally distributed (since a normal distribution divided by a constant is still normal). Consider two tightenings of $X\%$ and $Y\%$, where $X \geq Y$. Any training sample that must be retrained when tightening by $Y\%$ will also need to be retrained when tightening by $X\%$. Each point on the \texttt{Average} curve in Figure~\ref{fig:retrain} can be thought of as being the previous point plus a small $\delta$, where $\delta$ is the number of additional samples to retrain. The curve is thus the sum of these $\delta$s, and thus it approximates a Gaussian cumulative density function. \emph{In most cases,  \XCloud generates a set of alternative models that explore the performance vs. cost trade-off though stricter or more relaxed performance goals in under a minute}.

\subsection{Batch \& Online Query Scheduling}

  \begin{figure}
    \centering

    \includegraphics[width=0.45\textwidth]{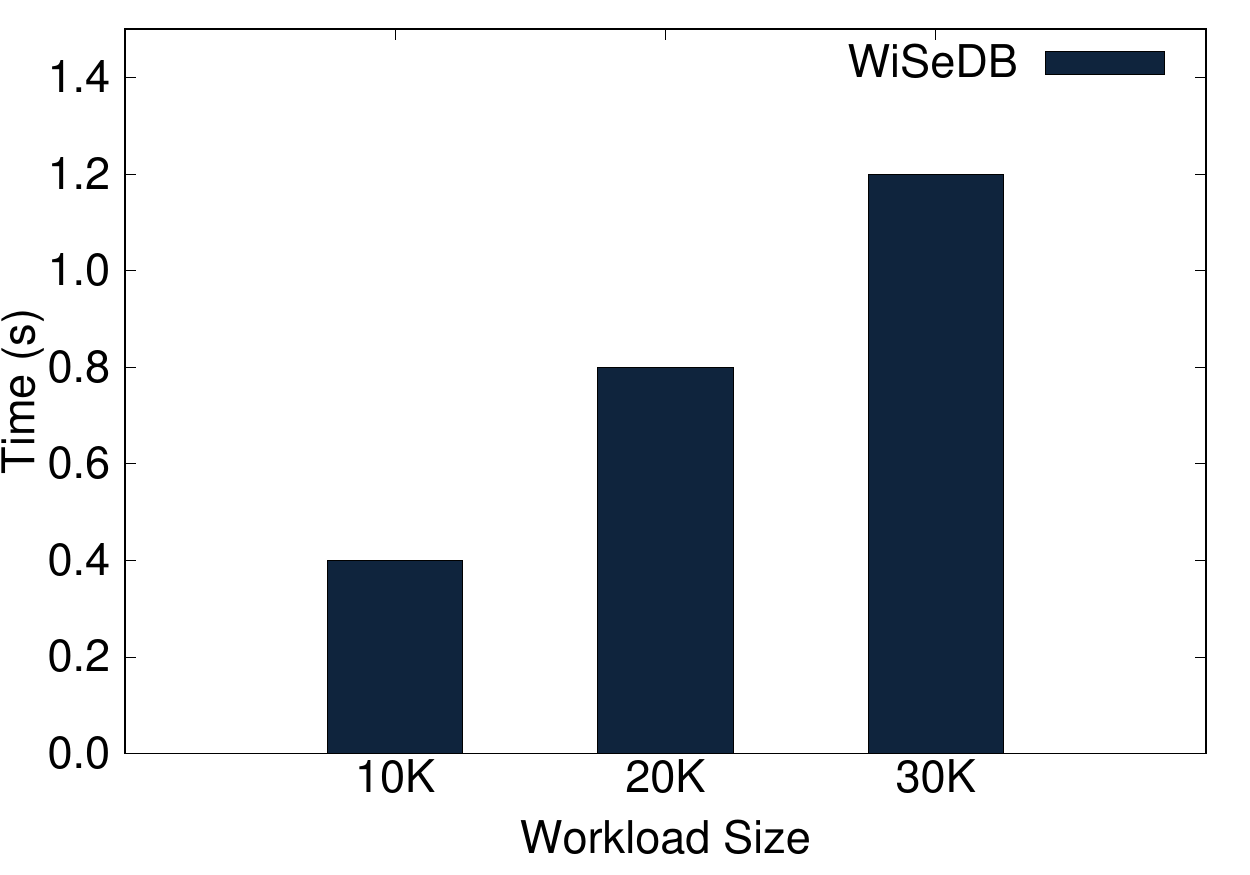}
    \caption{\small Scheduling overhead vs. batch size}
    \label{fig:runtime}
  \end{figure}
  \begin{figure}
    \centering

    \includegraphics[width=0.45\textwidth]{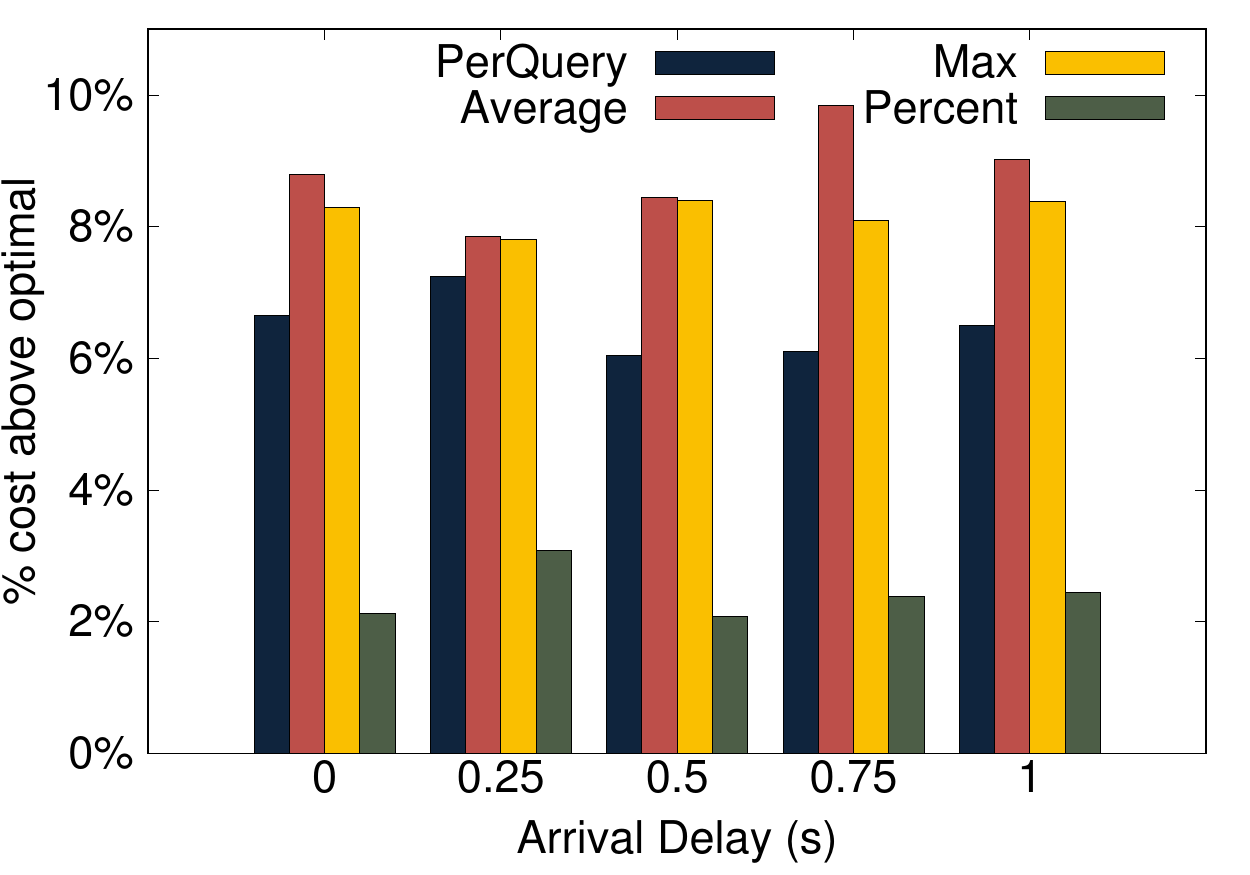}
    \caption{\small Effectiveness of online scheduling}
    \label{fig:arrival_rate}
  \end{figure}
  \begin{figure}
    \centering

    \includegraphics[width=0.45\textwidth]{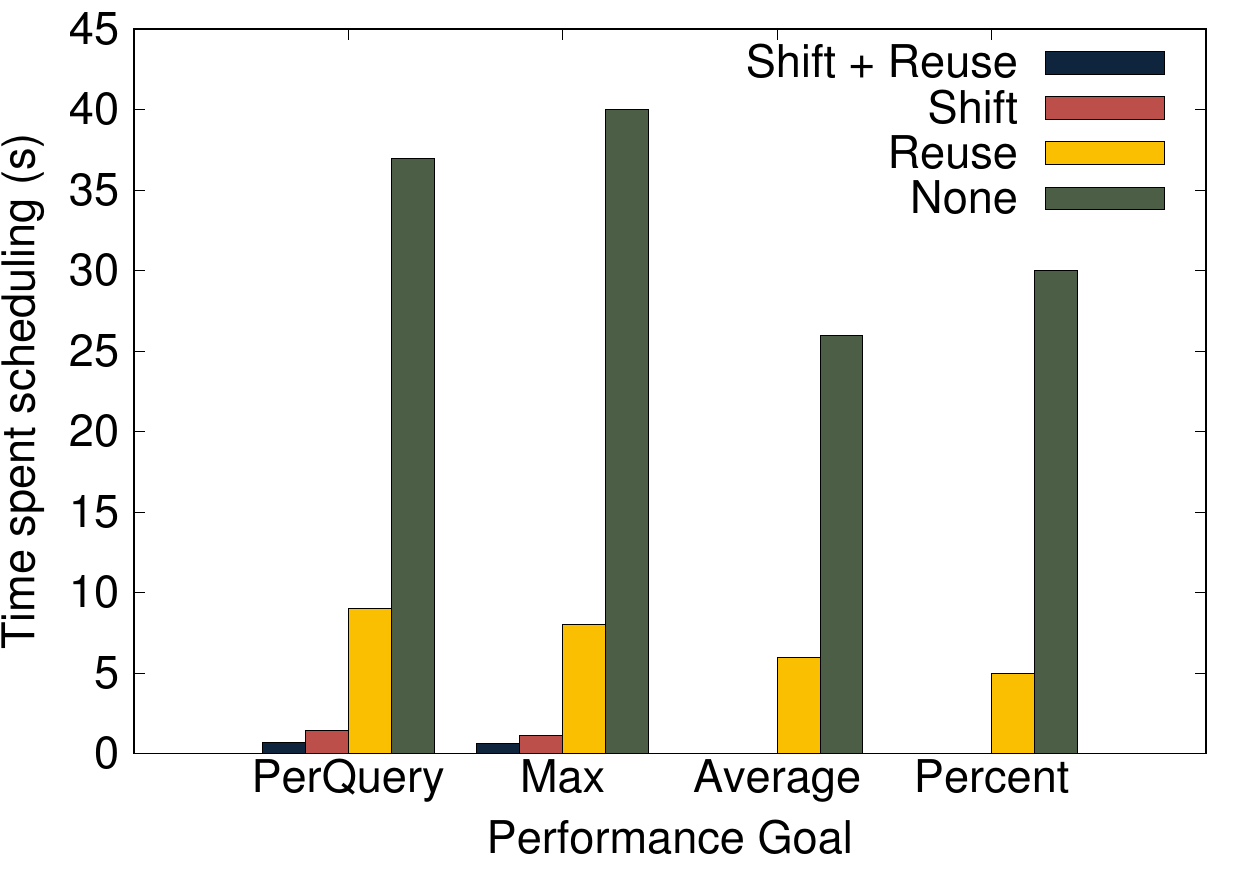}
    \caption{\small Average overhead for online scheduling}
    \label{fig:wait_time}
  \end{figure}

{\bf Batch Scheduling} {WiSeDB's models are used during runtime to generated schedules for incoming query batches. Section~\ref{s:effective} discussed the effectiveness of these schedules.  Figure~\ref{fig:runtime} shows the time required to generate schedules for workloads of various sizes. WiSeDB scales linearly with the number of queries and it can schedule up to 30,000 queries in under 1.5 seconds. WiSeDB's decision trees were usually shallow (height $h < 30$), so each decision to assign a query or to provision a new VM can be made quickly. The runtime complexity of generating schedules using a decision model is bounded by $O(h \times n)$, where $h$ is the height of the decision tree and $n$ is the number of queries in our batch. 

Systems or heuristics that must sort the queries in the batch beforehand, or systems that must scan over all provisioned VMs in order to make each query assignment or provisioning decision (e.g.,~\cite{sci_place, smartsla}), may not perform as well as \XCloud for very large workloads. Since there must be at least one query on each VM, the maximum number of times the tree needs to be parsed is $2n$, where $n$ is the number of queries in the batch. Each parse takes $O(h)$ time, so the runtime complexity of generating schedules using a decision model is bounded by $O(hn)$.

We note that the scalability of WiSeDB does not depend on the number of VMs (the number of VM to be rented is an output of our models).}
\emph{We thus conclude that WiSeDB's decision models scale linearly and can be used to schedule efficiently very large batch workloads.}

{\bf Online Scheduling} \XCloud also supports online scheduling by training a new model with additional query templates when a new query arrives. Section~\ref{sec:online} describes our approach as well as  two optimizations for reducing the frequency of model re-training. The model reuse optimization ({\tt Reuse}) can be applied to all four performance metrics. The linear shifting optimization ({\tt Shift}) can only be  used  for the \texttt{Max} and {\tt PerQuery} metrics. 

Figure~\ref{fig:arrival_rate} shows the performance (cost) of \XCloud compared to an optimal scheduler for various query arrival rates and performance metrics. For each metrics, we generate a set of $30$ queries and run them in a random order, varying the time between queries. More time between queries means that the scheduler can use fewer parallel VMs, thus reducing cost. The results demonstrate that \XCloud compares favorably with the optimal. In all cases, it generates schedules with costs that are within 10\% of the optimal. 

{Figure~\ref{fig:wait_time} shows the impact of our optimizations on average scheduling overhead, i.e., the average time a query waits before being assigned to a VM. Here, {\tt Shift} refers to using only the shifting optimization, {\tt Reuse} refers to using only the model reuse optimization, and {\tt Shift+Reuse} refers to using both optimizations. We compare these with {\tt None}, which retrains a new model at each query arrival.  We use a query arrival rate that is normally distributed with a mean of $\frac{1}{4}$ seconds and standard deviation of $\frac{1}{8}$. \cut{We also assume a $5\%$ error in the query latency prediction model.} If a query arrives before the last query was scheduled, it waits. 

Figure~\ref{fig:wait_time} shows that the average query wait time can be reduced to below a second for the {\tt PerQuery} and {\tt Max} performance goals using both the linear shifting and the model reuse optimization. The {\tt Average} and {\tt Percent} performance goals have substantially longer wait times, at around 5 seconds, but a 5 second delay represents only a $2\%$ slowdown for the average query.} \emph{Hence, our optimizations are very effective in reducing the query wait time, allowing \XCloud to efficiently reuse its generated models and offer online scheduling solutions in a timely manner}.

\subsection{Sensitivity Analysis}\label{s_sensitivity}

\begin{figure}
  \centering
  \includegraphics[width=0.45\textwidth]{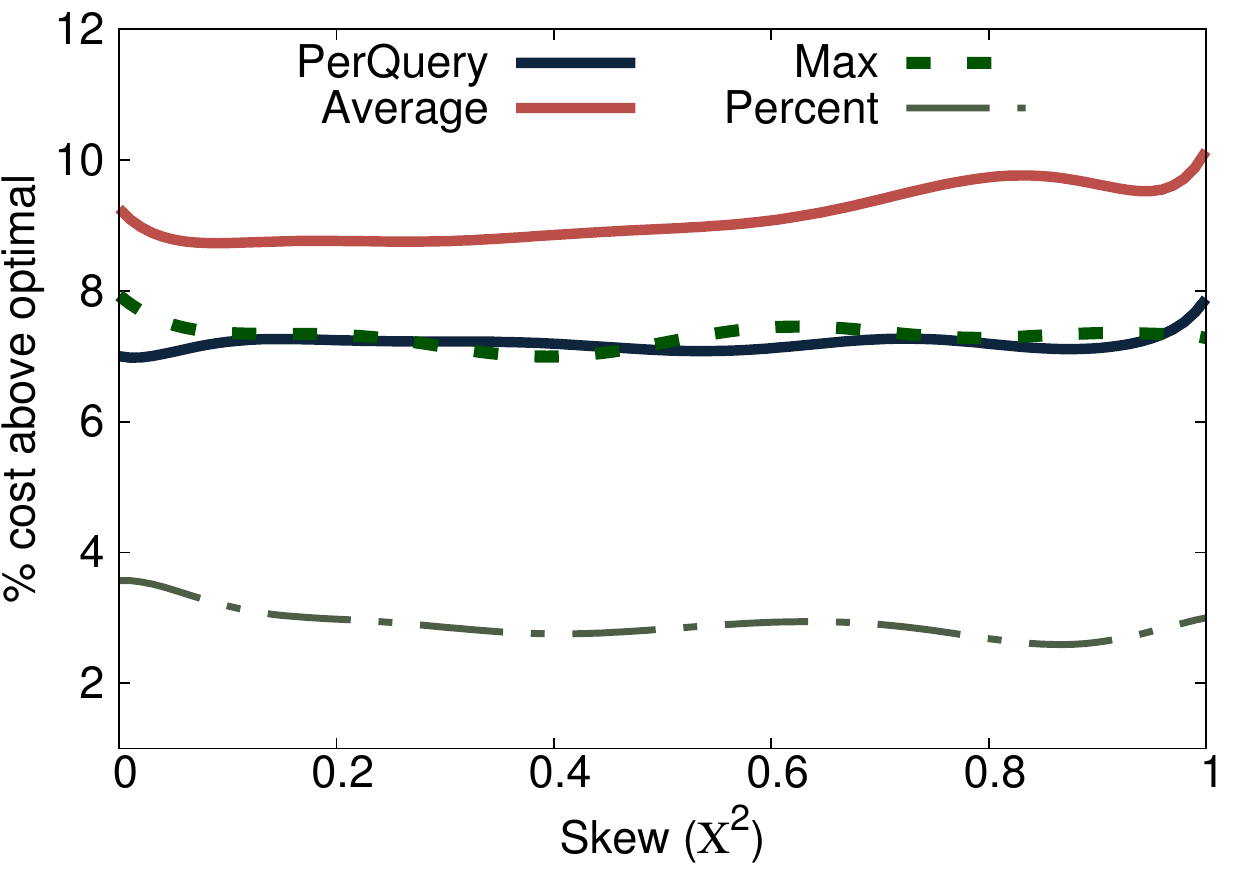}
  \caption{\small Sensitivity to skewed runtime workloads}
  \label{fig:avg_skew}
\end{figure}
\begin{figure}
  \centering
  \includegraphics[width=0.45\textwidth]{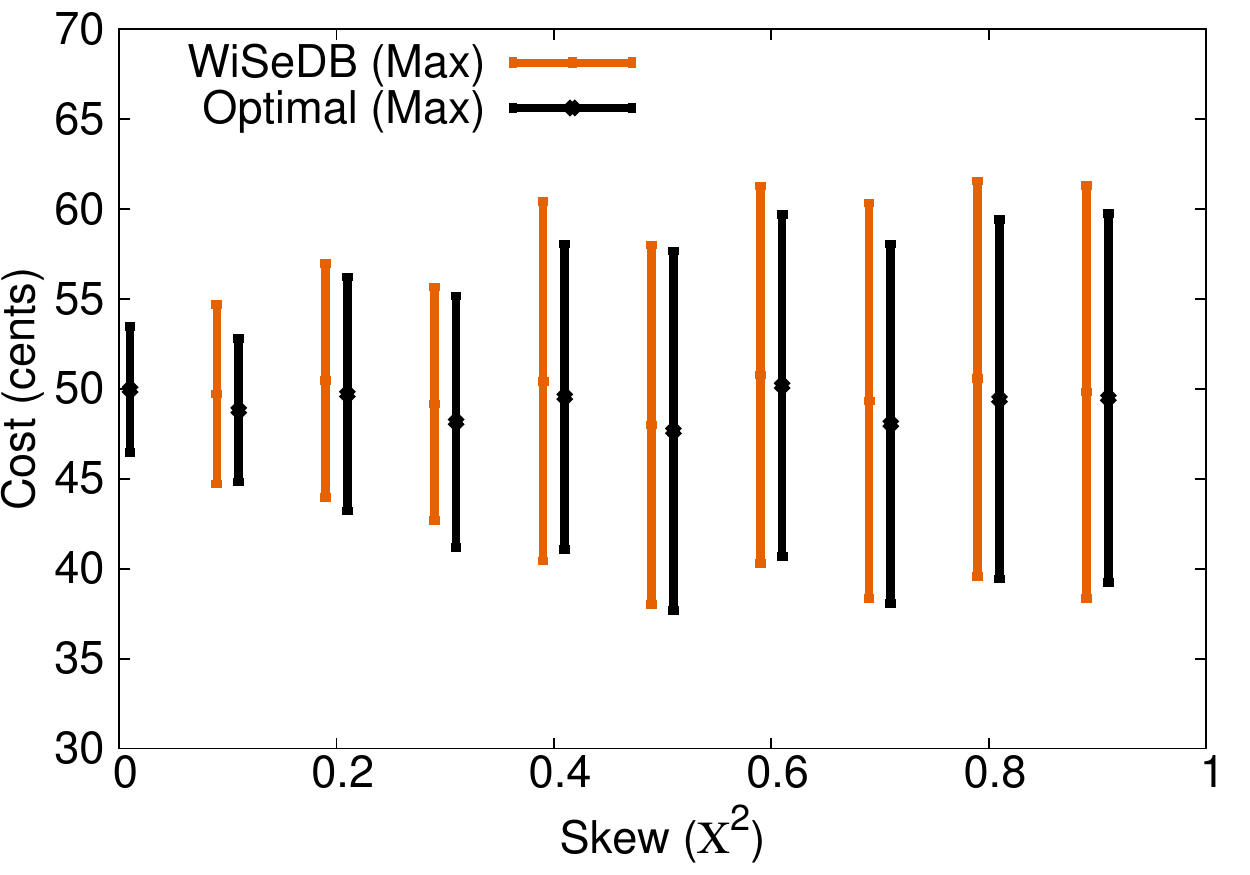}
  \caption{\small Workload skewness vs. cost range}
  \label{fig:range_skew}
\end{figure}
\begin{figure}
  \centering
  \includegraphics[width=0.45\textwidth]{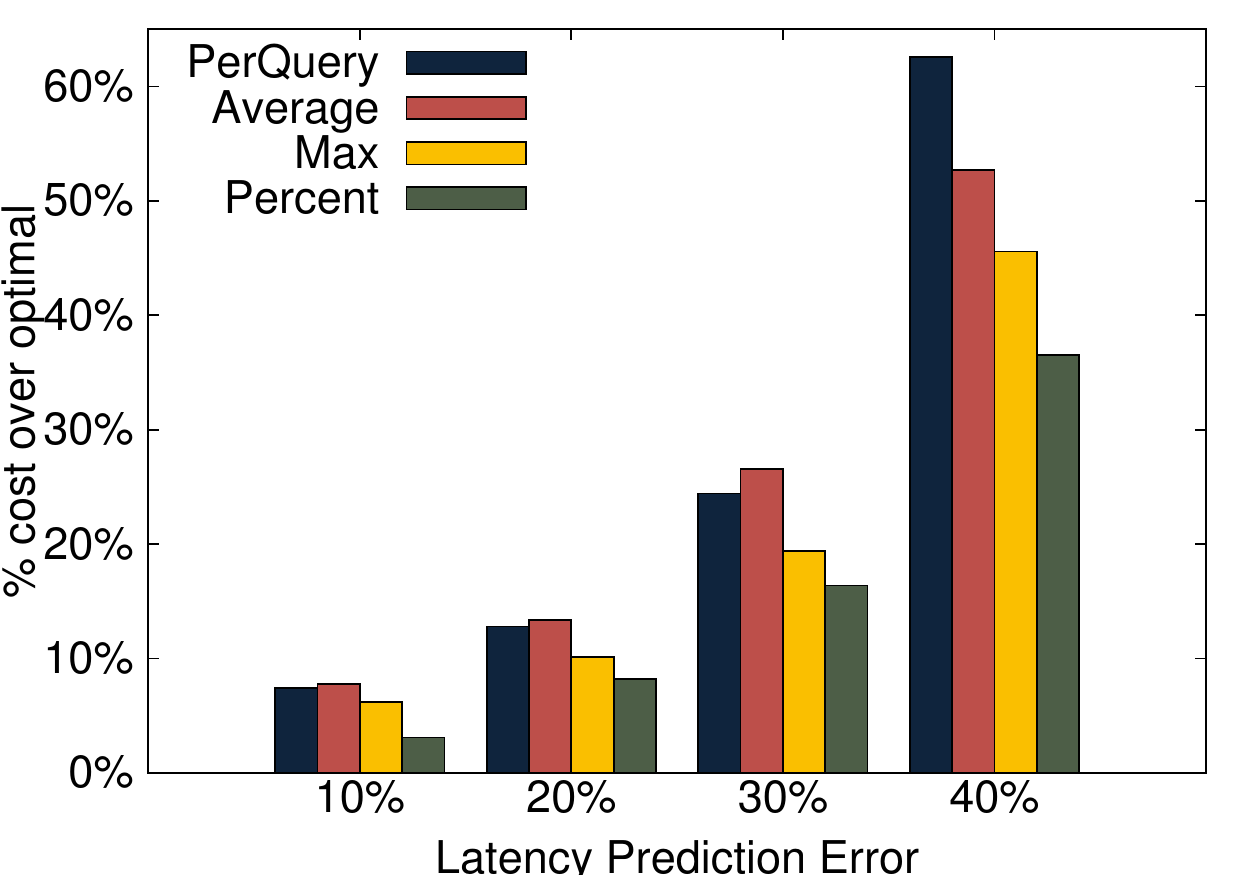}
  \caption{\small Optimality for varying cost errors}
  \label{fig:cm_sens}
\end{figure}

{\bf Skewed Workloads} We also experimented with scheduling workloads heavily skewed towards some query templates. Figure~\ref{fig:avg_skew} shows the average percent increase over the optimal cost for a workload with varying $\chi^2$ statistics~\cite{chi2}, which indicates the skewness factor. The $\chi^2$ statistic measures the likelihood that a distribution of queries was not drawn randomly: $0$ represents a uniform distribution of queries w.r.t. templates, and $1$ represents the most skewed distribution possible, i.e., a batch which includes only a single template (while the decision model is trained for workloads of 10 templates). The $\chi^2$ statistics were calculated with the null hypothesis that each query template would be equally represented. Thus, the value on the x-axis is the confidence with which that hypothesis can be rejected.
Even with highly skewed workloads consisting of almost exclusively a single query template ($\chi^2 \approx 1$), the average percent-increase over the optimal changes by less than $2\%$. 

To better understand these results, we used WiSeDB to schedule 1000 workloads (instead of our default 5 workloads) under different skewness factors for the {\tt Max} metric. Figure~\ref{fig:range_skew} shows both the average and the range of  the cost of these schedules. While the mean cost remains relatively constant across different $\chi^2$  values (as in Figure~\ref{fig:avg_skew}), the variance of the cost increases as skew increases. This is because a very skewed workload could contain a disproportionate number of cheap or expensive queries, whereas a more uniform workload will contain approximately equal numbers of each. \XCloud's decision models have variance approximately equal to that of an optimal scheduler. \emph{Hence, \XCloud's models perform effectively even in the presence of  skewed query workloads.}

\begin{figure}
\centering
\includegraphics[width=0.40\textwidth]{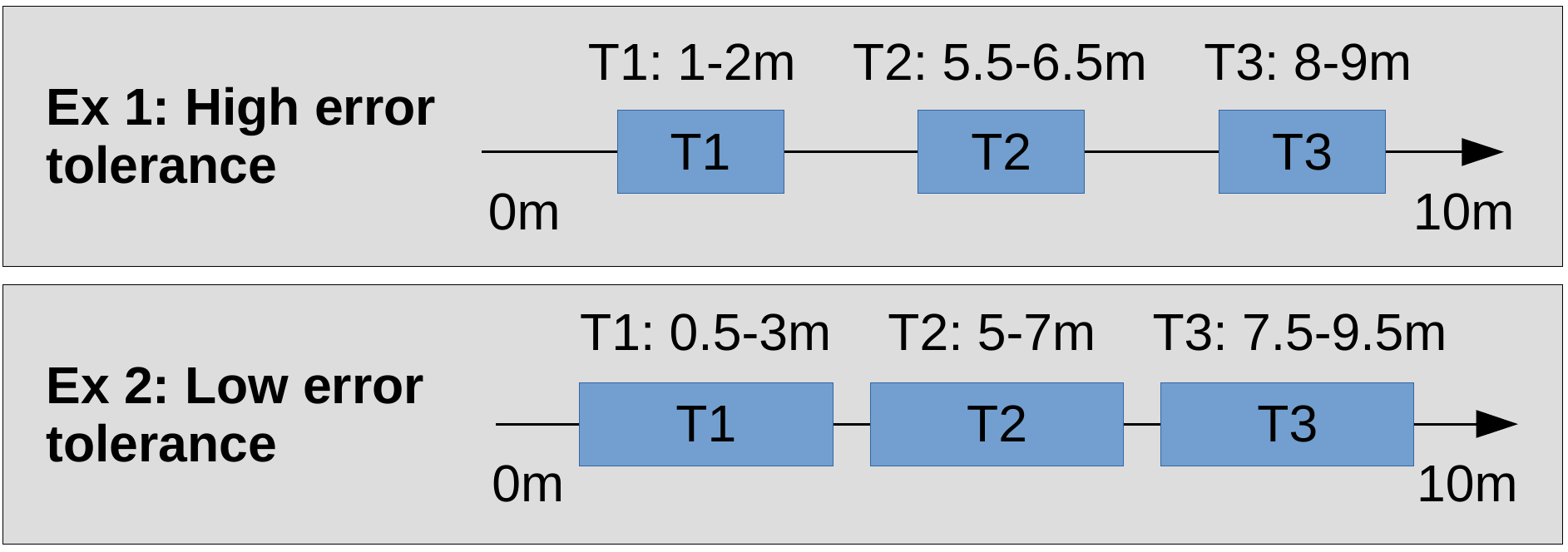}
\caption{\small{Example of templates with low and high error tolerance}}
\label{fig:error_tolerance}
\vspace{-5mm}
\end{figure}

{\bf Latency Prediction Accuracy} \XCloud relies on a query latency prediction model to estimate penalty costs. While existing query performance prediction models~\cite{contender, jennie_sigmod11} can be used, these often exhibit prediction errors, leading to incorrect estimations of the penalty costs. Naturally, we expect that the effectiveness of schedules produced by \XCloud to decrease as the cost model error increases, but we discovered that \XCloud is able to tolerate a certain level of prediction errors. \emph{In fact, the more distinguishable templates are with respect to their latency (see Figure~\ref{fig:error_tolerance}), the higher prediction error \XCloud can tolerate.} This is because \XCloud expects queries with similar latencies to be assigned to the same template. Therefore, the presence of latency prediction errors causes some queries to have ambiguous template membership, hindering \XCloud's ability to learn effective models.

Figure~\ref{fig:cm_sens} demonstrates this conclusion. Here, each $\sigma^2$ value refers to the cost model error (standard deviation) as a percentage of the actual query latency. \XCloud handles cost model errors less than $30\%$ very well. This is because, given our template definitions, the percentage of queries who are assigned to the wrong template at $30\%$ error, e.g. a query with actual latency similar to template $T_i$ is mistakenly assigned to template $T_j$, is $14\%$. At $40\%$ error, this percentage rises to $67\%$, leading to poor scheduling decisions.

%% file: related.tex
\section{Related Work}\label{s:related}
Many research efforts address various aspects of workload management in cloud databases. 
iCBS~\cite{icbs} offers a generalized profit-aware heuristic for ordering queries, but the algorithm considers assignments to a single VM, and performance goals are limited to piecewise-linear functions (which cannot express percentile performance metrics). 
In~\cite{sla-tree}, they propose a data structure to support profit-oriented workload allocation decisions, including scheduling and provisioning. However, their work supports only step-wise SLAs, which cannot express average or percentile goals.
In~\cite{pmax,slos,cloudoptimizer,delphi_pythia}, they consider the problem of mapping each tenant (customer) to a single physical machine to meet performance goals, but ordering and scheduling queries is left to the application. 
SmartSLA~\cite{smartsla} offers a dynamic resource allocation approach for multi-tenant databases. 
\XCloud supports a wider range of metrics than the above systems and its decision models offers holistic solutions that indicate the VMs to provision, query assignments, and query execution order. {\XCloud also \emph{learns} decision models for various SLA types, as opposed to utilizing a hand-written, human-tailored heuristic. This brings an advantage of increased flexibility (changing performance goals without reimplementation) at the cost of some training overhead.}


In ~\cite{q-cop, activesla}, they propose admission control solutions that reject queries that might cause an SLA violation at runtime, whereas our system seeks to minimize the cost of scheduling every query and to inform the user of performance/cost trade-offs. {Even if a query cannot be processed profitably, \XCloud will still attempt to place it in a cost-minimal way.} ~\cite{pslas} proposes multiple SLAs with different prices for various query workloads, but leaves query scheduling up to the application and supports only per-query latency SLAs, whereas we allow applications to define their own query \emph{and} workload-level performance metrics.  

In~\cite{sqlvm}, they propose monitoring mechanism for resource-level SLAs, and in~\cite{bazaar}, they propose an approach for translating query-level performance goals to resource requirements, but both assume only a single performance metric  and leave query scheduling up to the application. {\XCloud takes a query-centric as opposed to a resource-centric approach, assuming that a latency prediction model can correctly account of resource oscillations and query affinity.}
~\cite{opennebula} proposes an end-to-end cloud infrastructure management tool, but it supports only constraints on the maximum query latency within a workload and uses a simple heuristic for task scheduling. \XCloud learns specific heuristics for application-specific performance goal by training on optimal schedules for representative workload samples. 
In~\cite{sci_place}, they use a hypergraph partitioning approach to schedule  tasks expressed as directed acyclic graphs on cloud infrastructures. {While \XCloud contains no notion of query dependency, ~\cite{sci_place}  does not consider performance goals of any type, nor provisioning  additional resources.}
Finally, ~\cite{hadoop_place} proposes a system for scheduling Hadoop jobs on clouds which takes resource provisioning and deadlines into account. However, it considers only simple per-task deadlines and is not easily generalized beyond Hadoop.

%% file: conclusion.tex
\section{conclusions}
\label{s:conclusions}

{This work introduces \XCloud, a workload management advisor for cloud databases. To the best of our knowledge, \XCloud is the first system to address workload management in an \emph{holistic} fashion, handling the tasks of resource provisioning, query placement, and query scheduling for a broad range of performance metrics.} 
{\XCloud leverages machine learning techniques to learn decision models for guiding the low-cost execution of incoming queries under application-defined performance goals. We have shown that these decision models can efficiently and effectively schedule a wide range of workloads. These  models can be quickly adapted to enable exploration of the performance vs. cost trade-offs inherent in cloud computing, as well as provide online query scheduling with little overhead. Our experiments demonstrate that \XCloud can gracefully adapt to errors in cost prediction models, take advantage of multiple VM types, process skewed workloads, and outperform several well-known heuristics with small training overhead.}

{We have a full research agenda moving forward. We are currently investigating alternative features for characterizing the optimal assignment decision as well as alternative learning techniques (e.g., neural networks, reinforcement learning) for the workload management problem. We are also looking into multi-metric performance goals that combine workload and query level constraints, as well as dynamic constraints that change based on some external variable, e.g. time of day.}